\documentclass[aps, reprint,groupedaddress,longbibliography,superscriptaddress,nofootinbib]{revtex4-2}
\usepackage{graphicx, tabularx, longtable, nicefrac, amsmath,enumerate, supertabular, eurosym}
\usepackage{hyperref,xcolor}
\hypersetup{breaklinks=true, colorlinks = true, citecolor = blue!30!black, linkcolor = blue!30!black}
\usepackage{booktabs}

\newcommand{\env}[1]{\texttt{#1}}
\usepackage[utf8]{inputenc}
\usepackage[T1]{fontenc}
\usepackage{tikz}
\usepackage{pgfplots}
\pgfplotsset{compat=1.8}
\usepgfplotslibrary{polar}
\definecolor{bblack}{HTML}{363033}
\usepackage{xcolor, booktabs}
\definecolor{grun}{rgb}{0.0, 0.5, 0.0}
\definecolor{amber}{rgb}{1, 0.49, 0.0}
\definecolor{alizarin}{rgb}{0.82, 0.1, 0.26}
\definecolor{dblau}{RGB}{21,50,104}
\definecolor{hblau}{RGB}{132,191,234}
\definecolor{chamois}{RGB}{243,226,216}
\definecolor{altweis}{RGB}{246,244,240}
\definecolor{dgrau}{RGB}{0,101,141}
\definecolor{hgrau}{RGB}{87,87,86}
\usetikzlibrary{patterns, shapes,arrows,positioning, shadows,fit,backgrounds}
\usepackage{setspace}
\usepackage{tabu}
\usepackage{multirow}
\usepackage{xcolor}
\usepackage{calc}
\usepackage{amsmath,amsthm,amssymb,euscript} 
\usepackage{enumerate,graphicx,setspace}
\usepackage{times}
\usepackage{longtable}
\usepackage{url}
\usepackage{enumitem}
\usepackage{comment}
\usepackage{endnotes}
\usepackage{pgf-pie}
\usepackage{diagbox}
\usepackage{afterpage}
\usepackage{placeins}

\begin{document}

\title{From enrollment to exams: Perceived stress dynamics among first-year physics students}

\author{Simon Zacharias Lahme}
\email{simon.lahme@uni-goettingen.de}
\affiliation{Faculty of Physics, Physics Education Research, University of Göttingen, Friedrich-Hund-Platz 1, 37077 Göttingen, Germany}

\author{Jasper Ole Cirkel}
\affiliation{Faculty of Physics, Physics Education Research, University of Göttingen, Friedrich-Hund-Platz 1, 37077 Göttingen, Germany}

\author{Larissa Hahn}
\affiliation{Faculty of Physics, Physics Education Research, University of Göttingen, Friedrich-Hund-Platz 1, 37077 Göttingen, Germany}

\author{Julia Hofmann}
\affiliation{Faculty of Physics, Physics Education Research, University of Göttingen, Friedrich-Hund-Platz 1, 37077 Göttingen, Germany}

\author{Josefine Neuhaus}
\affiliation{Faculty of Physics, Physics Education Research, University of Göttingen, Friedrich-Hund-Platz 1, 37077 Göttingen, Germany}

\author{Susanne Schneider}
\affiliation{Faculty of Physics, Physics Education Research, University of Göttingen, Friedrich-Hund-Platz 1, 37077 Göttingen, Germany}

\author{Pascal Klein}
\affiliation{Faculty of Physics, Physics Education Research, University of Göttingen, Friedrich-Hund-Platz 1, 37077 Göttingen, Germany}

\date{\today}

\begin{abstract}
The current dropout rate in physics studies in Germany is about 60\%, with the majority of dropouts occurring in the first year. Consequently, the physics study entry phase poses a significant challenge for many students. Students' stress perceptions can provide more profound insights into the processes and challenges during that period. In a panel study featuring 67 measuring points involving up to 128 participants at each point, we investigated students' stress perceptions with the Perceived Stress Questionnaire (PSQ), identified underlying sources of stress, and assessed self-estimated workloads across two different cohorts. This examination occurred almost every week during the first semester, and for one cohort also in the second semester, yielding a total of 3,241 PSQ data points and 5,823 stressors. The PSQ data indicate a consistent stress trajectory across all three groups studied that is characterized by significant dynamics between measuring points, spanning from $M=20.1, SD=15.9$ to $M=63.6, SD=13.4$ on a scale from 0 to 100. Stress levels rise in the first weeks of the lecture, followed by stable, elevated stress levels until the exams and a relaxation phase afterward during the lecture-free time and Christmas vacation. In the first half of the lecture period, students primarily indicated the weekly exercise sheets, the physics lab course, and math courses as stressors; later on, preparation for exams and the exams themselves emerged as the most important stressors. Together with the students' self-estimated workloads that correlate with the PSQ scores, we can create a coherent picture of stress perceptions among first-year physics students, which builds the basis for supportive measures and interventions. 
\end{abstract}

\maketitle

\section{Motivation: High dropout in physics studies}

In the academic year 2023, approximately 11,100 students were enrolled in undergraduate physics programs at German universities \cite{Duchs.2023}. This figure has returned to the levels observed in the academic year 2011, marking a significant decrease from the intervening years, during which enrollment numbers ranged between 14,000 and 16,000 new students annually. Of these students, about 39\% are so-called park students, who do not attend any courses in person, since they do not intend to study physics, but only enroll for appearance's sake, for example, to wait for another study or training place, or to take advantage of all the social benefits of being a student. Because of this effect, only about 6,800 students actually started their studies in the recent academic year \cite{Duchs.2023}. Given the often-discussed shortage of skilled workers, especially in fields related to the natural sciences, a problem exacerbated by demographic transition, the number of students enrolling is considered to be particularly low. This lack of enrollment in physics is even more dramatic given the high dropout rate among physics students. The exact rate is difficult to determine and depends on the measurement used. 

A commonly referenced metric, established by the German Centre for Higher Education Research and Science Studies \cite{Heublein.2022b}, assesses the proportion of a specific reference cohort of students who complete their degree programs, regardless of whether they graduate in their original subjects or from their initial universities. According to the latest data, which are based on a comparison of the 2020 cohort of graduates with students who began their studies in 2016/17 or before, 60\% of all students enrolled in a physics or geosciences bachelor's degree study program at German universities did not receive any degree. This is currently the highest rate compared to all other study programs, and it has steadily been one of the highest dropout rates in recent years, with an upward trend from 39\% among students who started in 2006/07 to 49\% in 2014/15 and 60\% in 2016/17 \cite{Heublein.2017,Heublein.2022b}. Another method to estimate the dropout rate involves comparing the actual number of physics bachelor's degree graduates (2,660) in 2023,\footnote{For comparison, in the US, which has around four times more inhabitants than Germany, 9,031 students graduated from a physics bachelor's degree program in 2020 \cite{Nicholson.2022}; in the UK, which has around 20\% fewer inhabitants than Germany, 3,675 students graduated from undergraduate physics courses in 2015 \cite{InstituteofPhysics.2017}.} which is consistent with figures from previous years \cite{Duchs.2023}, with the aforementioned enrollment numbers. This yields a success rate of less than 25\%.

For physics teacher training programs, these problems are even more serious as out of around 2,000 students enrolled annually, only around 300 complete a bachelor's degree, and slightly fewer a consecutive master's degree \cite{Duchs.2023}. Forecasts on the recruitment needs for new teachers show that the number of new graduates is far below the actual demand \cite{Klemm.2020}.

Low numbers of natural sciences students and high dropout are also international issues. Natural sciences, mathematics, and statistics have the lowest share of graduates in OECD countries compared to all other fields \cite{OECD.2020}, so physics is a very narrow field of study \cite{OECD.2023}. Data from the late 1990s and early 2000s show that high dropout in physics has already been an issue for a long time in various European countries \cite{Troendle.2004}.

In about half of these cases, dropout has already occurred in the first year of study \cite{Neugebauer.2021}, highlighting the necessity of focusing research efforts on understanding the underlying processes in this physics study entry phase. Here, stress perception can be considered a pivotal element for better understanding students' experiences that could lead to them dropping out (e.g., difficulties, challenges, circumstances). Therefore, this paper examines students' stress perceptions in that particular phase. 

\section{State of research and research questions}

\subsection{Research on the physics study entry phase}\label{Germanintroductory}

Study dropout rates and underlying causes have already been investigated and discussed in Germany for more than 40 years \cite{Reissert.1983}. Besides descriptive trend studies for all subjects like the previously cited study by Ref.~\cite{Heublein.2022b}, there has also been subject-specific research onscience and physics study programs in particular. In this context, three strains of research can be identified: research on risk factors, research on success factors, and efforts to innovate the study entry phase.

The first strain on risk factors investigates the extent of dropout, underlying processes, and reasons. Ref.~\cite{Albrecht.2011}, for example, has examined risk factors for a successful start in a physics study program. The study is based on work by Ref.~\cite{Thiel.2008}, who developed a model for academic success based on students' individual study and learning behavior that depends on personal entry preconditions (e.g., prior knowledge from school, study choice motives, or socio-demographic background), study conditions (e.g., quality of university teaching, the structure of the study program, or organization), and more general context conditions (e.g., employment, illness, or family situation). Ref.~\cite{Albrecht.2011} considered aspects like university entrance qualification grades, supervision, support, insufficient information, and interest in the subject and identified content requirements as the most salient reason for exmatriculation from a physics study program. 

The second strain investigates factual and affective aspects of the physics study entry phase and their impact on academic success. Examples are students' prior knowledge in mathematics and physics \cite{Krause.1981,Buschhuter.2016,Muller.2018} and their overall study ability \cite{Sorge.2016}, their acquisition of factual knowledge and problem-solving competencies during the introductory phase \cite{Woitkowski.2015,Woitkowski.2019}, students' views of the nature of science \cite{Woitkowski.2021}, or their sense of belonging to the physics community and university \cite{Feser.2023d}. Similar research has also been conducted in other countries, focusing on topics like the importance of math prerequisites \cite{Burkholder.2021} and the sense of belonging in the US \cite{Edwards.2022,Li.2023} or the emotions of physics students in Australia \cite{Bhansali.2019} and Finland \cite{Timonen.2022}.

The third strain, focusing on innovation, is about the improvement of the physics study entry phase to better support the teaching and learning process and facilitate overall academic success. Typical innovations are the implementation of extracurricular measures to tackle or compensate for discrepancies between actual and expected student performance. For example, learning centers have been established to provide supportive learning services in addition to the regular curriculum (e.g., \cite{Haak.2017}). Pre-courses have been introduced as on-campus courses and online learning units so that students can repeat relevant school mathematics, bridging the gap between school and university mathematics \cite{Bausch.2014,Gerdes.2022,Gerdes.2022b,Goll.2024,Bach.2024}. Additionally, efforts have been dedicated to the intracurricular enhancement of physics study programs. Examples include the implementation of classroom exercises, complexity-graded tasks, supportive learning materials for subject-specific problem-solving \cite{Bauer.2023,Woitkowski.2020,Abbas.2023}, or smartphone-based experimental tasks \cite{Staacks.2022,Hutz.2019,Kaps.2022,Klein.2014b}. A review of innovations' impact on learning in undergraduate science courses is available in Ref.~\cite{RuizPrimo.2011}.

\subsection{Perceived stress and stressors in university education}\label{stress}

\subsubsection{Definition of stress}
Following the first strain of research, i.e., investigating risk factors for study dropout, the exploration can extend to students' stress perceptions and the underlying causes of stress. Psychological stress can be defined as "a particular relationship between the person and the environment that is appraised by the person as taxing or exceeding his or her resources and endangering his or her well-being" \cite[p.19]{Lazarus.1984}. Stress perception is based on a cognitive appraisal process "of categorizing an encounter, and its various facets, with respect to its significance for well-being", which is of a "largely evaluative" and continuous nature [p.31]. The appraisal process takes into account both person and situation factors and particularly depends on the availability of strategies for coping, i.e., the "constantly changing cognitive and behavioral efforts to manage specific external and/or internal demands that are appraised as taxing or exceeding the resources of the person" [p.141]. As stress perception is a very individual interplay between a person and their environment, the same stressor, i.e., "a subset of environmental conditions that are likely to be appraised as demanding and to have implications for a person's well-being" \cite[p.30]{Moos.1990}, can be perceived differently by different people.

\subsubsection{Measuring stress perceptions: Perceived Stress Questionnaire}

A popular instrument for measuring stress perception is the Perceived Stress Questionnaire (PSQ), originally conceived for the investigation of psychosomatic patients \cite{Levenstein.1993}. It has been translated into various languages, including German \cite{Fliege.2001}, Spanish \cite{SanzCarrillo.2002}, and Chinese \cite{Meng.2020,Jiang.2023}. The language versions differ in the number of items and subscales. The German version comprises twenty items (statements), and a four-point Likert-type rating scale to assess the level of agreement with each statement, viz. \textit{almost never} (1), \textit{sometimes} (2), \textit{often} (3), and \textit{most of the time} (4). They form four subscales with five items each, which describe the perceived \textit{worries}, \textit{tension}, and \textit{joy} that seem to reflect the internal stress reaction of the individual, as well as \textit{demands} that seem to refer to the perception/appraisal of external stressors \cite{Fliege.2001}. Worries, tension, and demands are treated as increasing while joy is treated as decreasing the total perceived stress score that is derived as the mean across all items after partial inversion of items. The score is scaled linearly from 0 (min.) to 100 (max.).

In a broader body of research, the PSQ has been used to study students in a variety of disciplines (cf. Table~\ref{tab:comparison}). It has been used in the fields of medicine, nursing, and health, particularly to validate the different versions of the PSQ, to investigate education students, or to obtain an overview of stress perception at universities, e.g., during the COVID-19 pandemic, by investigating students in various subjects and years of study. Data were collected on different time scales from a few days to several months and always in one measuring period, except for Ref.~\cite{Obermeier.2022}, who measured three times at the beginning, middle, and end of a semester. The sample sizes, means, and standard deviations vary considerably.

\begin{table*}[htb]
\caption{Overview of the literature providing comparative data from the Perceived Stress Questionnaire (PSQ) for students. For each reference, the version of the instrument used is indicated, i.e., the number of items (13, 20 or 30), the language (Chinese (C), English (E), German (G), Spain (S)), and whether a four- or six-point rating scale was used. Where identifiable, a description of the students studied and the measuring points are given, together with the number of students (N) and their mean total perceived stress score (M) including standard deviation (SD) and standard error (SE). For better comparability, all perceived stress scores are presented on a scale from 0 to 100, for which some data had to be transformed. For example, sometimes total stress scores had to be determined based on subscale scores. In this case, the average of the subscale scores and the highest standard deviation were used for approximation.}
\begin{ruledtabular}
\footnotesize
\begin{tabular}{p{.07\textwidth}p{.1\textwidth}p{.31\textwidth}p{.27\textwidth}p{.05\textwidth}p{.05\textwidth}p{.05\textwidth}p{.05\textwidth}}
Reference&Instrument&Investigated students&Measuring point&N&M&SD&SE\\\hline
\cite{Obermeier.2022}&PSQ-20-G-6&Across all subjects \& years of study including PhD, University of Nürnberg-Erlangen, Germany&begin of summer semester 20&2795&42.10&22.80&0.43\\
&&&mid of summer semester 20&2795&48.60&23.00&0.44\\
&&&end of summer semester 20&2795&50.50&23.40&0.44\\
\cite{Obermeier.2021}&PSQ-20-G-6&Education (primary \& secondary school), University of Nürnberg-Erlangen, Germany, various years of study&begin of summer semester 20&51&39.00&27.80&3.89\\
&&&mid of summer semester 20&51&49.90&25.40&3.56\\
&&&end of summer semester 20&51&51.30&27.00&3.78\\
\cite{Fliege.2001}&PSQ-20-G-4&Medicine, Germany, 4th or 5th clinical semester&&246&34&16&1.02\\
\cite{Buttner.2013}&PSQ-20-G-4&Across all subjects \& years of study, several universities in Germany&12/11-01/12&2435&52&19&0.39\\
\cite{Heinen.2017}&PSQ-20-G-4&Medicine, University Medical Center Hamburg-Eppendorf, Germany, 1st year of study&01/14 during regular seminar, at the beginning of the second six-week module before the written semester exam in February&321&40&15&0.84\\
\cite{Goppert.2021}&PSQ-20-G-4&Education, part of an introductory psychology lecture, University of Bamberg, Germany, various years of study&one measurement in winter semester 18/19, summer semester 19, winter semester 19/20, \& summer semester 20&110&40.75&23.6&2.26\\
\cite{Moldt.2022}&PSQ-20-G-4&Medicine, part of a communication course, Germany&winter semester 20/21&136&49.53&10.64&0.91\\
\cite{LargoWight.2005}&PSQ-30-E-4&Health \& activity course, Mid-Atlantic University, US, undergraduate level&&232&62.8&12.7&0.83\\
\cite{vanderFeltzCornelis.2020}&PSQ-30-E-4&Across all subjects \& years of study, University of York \& Hull York Medical School, UK&05/20-06/20&788&51&20&0.71\\
\cite{MonteroMarin.2014,MonteroMarin.2014b}&PSQ-30-S-4&Dental, Universities of Uesca \& Santiago de Compostela, Spain, various years of study&05/11, two weeks before final exam phase&314&45&19&1.07\\
\cite{MartinezRubio.2023}&PSQ-24-S-4&Education, Catholic University of Valencia, Spain, undergraduate level&end of classes, 03/16-05/16&589&42&17&0.70\\
\cite{Meng.2020,Luo.2018}&PSQ-30-C-4&Medicine, University of Wuhan, China, post- \& undergraduate&12/16-01/17&122&40.2&13.3&1.20\\
&&Medicine, University of Ningbo, China, junior college&11/15-01/16&1453&39.9&13.8&0.36\\
\cite{Jiang.2023}&PSQ-13-C-4&Medicine, University of Hangzhou, China, various years of study&09/21-12/21&309&29.019&6.325&1.944\\
\end{tabular}
\end{ruledtabular}
\label{tab:comparison}
\end{table*}

\subsubsection{Further research on students' stress perceptions}

Beyond the use of the PSQ, there have been further studies about students' stress perceptions. Some of them do not explicitly focus on a specific subject \cite{Ortenburger.2013,Zhao.2023,Herbst.2016,Brandt.2023,Denovan.2019,Gro.2011}, while others address specific subjects like nutrition \cite{Pitt.2018}, teacher training students \cite{Bauer.2019b,Hahn.2021,Vogelsang.2021}, or chemistry \cite{Schwedler.2017}. Research interests, objectives, and methods vary widely in these studies, and this list is not exhaustive. Therefore, we only highlight three studies that are particularly relevant in our research context. 

Ref.~\cite{Ortenburger.2013} used the Perceived Stress Scale by Ref.~\cite{Cohen.1983} among German students over two weeks in late November 2011, leading to the observations that 59\% of the students felt nervous or stressed within the past two weeks and 31\% had the feeling that they could not control important aspects in their lives. No relevant differences between subjects were found, but women and students in more advanced semesters tended to perceive more stress. Stress was mostly associated with time and performance pressure (greater in earlier years of study), fear of the future and uncertainty (greater in later years of study), and excessive demands posed on students. In the study, students also rated eleven different areas of life (studies, work, family situation, financial situation, leisure, health, household, children, partnership, social contacts, and housing situation) identified by Ref.~\cite{Schroder.2007} according to which extent they contributed to their stress perceptions. The findings show that on average, each student perceived three areas of life as strong stressors. The most frequent stressors among all participants were the studies themselves (68\% of students) as well as financial situations, work, and leisure (around 40\% each).

Similarly, Ref.~\cite{Gro.2011} investigated the relevance of ten predefined stressors as perceived by 243 students of various subjects and years of study at a German university. The responses revealed that the general conditions of the universities, like rooms, facilities, or number of participants in courses; the requirements regarding the course of studies and timetable in major subjects; individual financial situations, and working styles as a fit between habits and external requirements were the most important stressors. Requirements in minor subjects, conflicts between studies and private interests, and part-time jobs were perceived as being of medium importance. The need for active participation in courses, troubles with self-concepts of ability, and family obligations were considered unimportant stressors for most students.

Ref.~\cite{Schwedler.2017} examined the stress of 178 first-semester chemistry students at two German universities, also following the definition of stress by Ref.~\cite{Lazarus.1984}. Monthly, students rated on a five-point Likert scale a single item by Ref.~\cite{Schulz.2004}: to what extent they felt that they could meet the demands of their study program in the same week. Furthermore, they could state reasons for their feeling of overload in an open text field. The quantitative data show that 38\% of all students perceived a moderate and 26\% a high or very high mismatch (similar to the level of stress) during their first year of studying chemistry. The analysis of the qualitative data leads to four areas of stressors. The most prominent area is built by cognitive-factual stressors, which were split into qualitative (i.e., level of difficulty), quantitative (i.e., workload), and not further specifiable stressors, as well as responses regarding exam failure. Furthermore, cognitive-organizational, physical (including sickness), and social stressors were identified but were less common. While qualitative stressors dominated the first half of the first semester, the second half of the semester was dominated by quantitative stressors due to lab courses and exam preparation. The math and lab courses and particularly the weekly exercises and self-study at home were the most relevant aspects linked to these quantitative and qualitative stressors, while exercises and tutorials were rarely mentioned.

\subsubsection{Stress and the role of academic workload}

An accompanying factor that is sometimes investigated together with students' stress perceptions is academic workload. For instance, Ref.~\cite{Buttner.2013} utilized the PSQ alongside an item by Ref.~\cite{Gusy.2010}, in which students were asked to report the number of hours they spent in an average lecture week on attending courses, engaging in course-related activities, and working to finance their studies. The analysis shows that self-estimated workload explained a significant part of the variance in the PSQ sub-dimensions of tension and demands, showing that workload is an interesting variable to consider together with the PSQ. The average workload reported in that study was comparable to a typical full-time job as around half of the students reported a workload between thirty-one and fifty hours per week, with extreme values even beyond seventy hours.

Conversely, an extensive time budget survey \cite{Schulmeister.2011b} across various subjects and German universities required students to detail their daily timetables, providing in-depth information on activities regarding their studies and estimates of private time, part-time jobs, vacation, illness, etc. The findings show that students spent significantly less time on their studies than one would expect. Based on the EU-wide ETCS system, students should spend around 1,800 hours per year on their studies, i.e., forty-five weeks of forty hours (and seven weeks of holidays), equal to on average 34.6 hours per week or 4.9 hours per day. However, at no university and in no study program investigated by Ref.~\cite{Schulmeister.2011b} was this 4.9 hour per day limit exceeded in any month of the semester (besides one minor exception). Usually, this limit is even significantly undercut (with the lowest value at 1.4 hours per day), leading, on average, to a twenty-three-hour week, significantly below the 34.6-hour estimate. The authors point out the comparability of their own findings with those from two American studies \cite{Babcock.2010,Babcock.2010b} showing that the workload of four-year college students decreased from forty hours per week in 1961 to around twenty-seven hours per week in 2003 \cite{Babcock.2011}.

\subsection{Research questions}\label{rq}

The state of the research indicates a considerable interest in examining students' stress perceptions across various subjects and countries. Yet, it also highlights a significant gap: despite physics study programs having the highest dropout rate among all study programs in Germany, there appears to be no study specifically focused on the stress perceptions of physics students. Furthermore, many studies have only one or very few measuring point(s) that do not allow for a high resolution of possibly time-dependent stress perception and stressors in the first year of study. However, the PSQ seems to be an established and frequently validated instrument to investigate stress perception of university students. Facing the need to better understand first-year physics students' stress perceptions and stressors, we pursue the following research questions:

RQ1a: \textit{How does the perceived stress of physics students (measured by the PSQ) evolve during the first year of their study?}
This research question deals with the temporal dependency of the stress perceptions of physics students in their first year of study. An answer to this question will provide quantitative data on which phases of the study entry phase are perceived as more or less stressful by students, as well as an overall assessment of their stress levels.

RQ1b: \textit{How does the self-estimated workload of students evolve during the first year of their study?}
This research question is analogous and complementary to the previous one. It deals with students' self-assessed weekly study-related workload as another indicator that could be related to stress perceptions. This will provide another variable in the context of understanding students' stress perceptions and may allow for a more coherent and complementary picture in this regard.

RQ2: \textit{What are the most salient stressors contributing to physics students’ perceived stress, and how does their indication shift during the first year of their studies?}
The third question goes beyond the quantitative data by looking at the stressors that contribute to perceived stress from students' point of view, and as above, the temporal dependency of these stressors over the semester is investigated. An answer to this question provides deep insight into the (subject-specific) causes of stress perception. This will be an important basis for future supportive measures and systemic improvements.

In addition, we investigate whether there are differences between different cohorts and between first- and second-semester students. This will show whether stress perception depends on the cohort and associated circumstances (e.g., at the university) and whether the first semester differs from the second semester, in which students are already more familiar with the university and their study program.

\section{Methods}

\subsection{Circumstances and mode of data collection and sample}

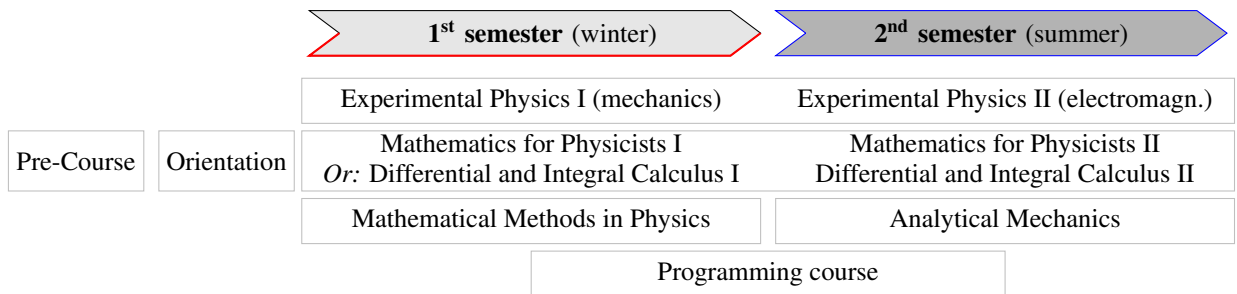
\begin{figure*}[htb]
    \centering
    \begin{tikzpicture}

  \draw[draw=black, fill=gray!20] (2.8,0)--(2.4,0.3)--(8.0,0.3)--(8.4,0)--(8.0,-0.3)--(2.4,-0.3)--(2.8,0);
  \draw[draw=red, thick](8.4,0)--(8.0,-0.3)--(2.4,-0.3)--(2.8,0);
  \node[ align=center, text width=14.5em] at (5.5,0){\textbf{1$^{\text{st}}$ semester} (winter)};
 \draw[draw=blue, fill=gray!60] (9.0,0)--(8.6,0.3)--(14.2,0.3)--(14.6,0)--(14.2,-0.3)--(8.6,-0.3)--(9.0,0) ;
 \node[ align=center, text width=14.5em] at (11.6,0){\textbf{2$^{\text{nd}}$ semester} (summer)};

  \draw[draw=gray!50] (2.3,-0.6) -- (14.7,-0.6) -- (14.7,-1.2) -- (2.3,-1.2) -- cycle;
  \draw[draw=gray!50] (2.3,-1.3) -- (14.7,-1.3) -- (14.7,-2.1) -- (2.3,-2.1) -- cycle;
  \draw[draw=gray!50] (2.3,-2.2) -- (8.4,-2.2) -- (8.4,-2.8) -- (2.3,-2.8) -- cycle;
  \draw[draw=gray!50] (14.7,-2.2) -- (8.6,-2.2) -- (8.6,-2.8) -- (14.7,-2.8) -- cycle;
  \draw[draw=gray!50] (5.35,-2.9) -- (11.65,-2.9) -- (11.65,-3.5) -- (5.35,-3.5) -- cycle;
  \draw[draw=gray!50] (0.4,-1.3) -- (2.2,-1.3) -- (2.2,-2.1) -- (0.4,-2.1) -- cycle;
  \draw[draw=gray!50] (-1.6,-1.3) -- (0.2,-1.3) -- (0.2,-2.1) -- (-1.6,-2.1) -- cycle;
  
  \node[align=center] at (5.35,-0.9){ Experimental Physics I (mechanics)};
  \node[align=center] at (5.35,-1.7){ Mathematics for Physicists I\\ \textit{Or:} Differential and Integral Calculus I};
  \node[align=center] at (11.65,-0.9){ Experimental Physics II (electromagn.)};
  \node[align=center] at (11.65,-1.7){ Mathematics for Physicists II\\ Differential and Integral Calculus II};
  \node[align=center] at (5.35,-2.5){ Mathematical Methods in Physics};
  \node[align=center] at (11.65,-2.5){ Analytical Mechanics};
  \node[align=center] at (8.5,-3.2){ Programming course};
  \node[align=center] at (1.3,-1.7){ Orientation};
  \node[align=center] at (-0.7,-1.7){ Pre-Course};
\end{tikzpicture}\vspace{-0.2cm}
    \caption{Structure of the first and second semester of the physics bachelor's study program at the University of Göttingen.}
    \label{fig:StudyStructure}
\end{figure*}

\begin{figure*}
\centering
\vspace{-0.5cm}
\begin{tikzpicture}
  \draw[ultra thick] (-2.7,1.3) node[below=3pt] {\color{black}\textbf{Group A1}} node[above=10pt] {};

  \draw[ultra thick] (9.2,1.3) node[below=3pt] {\color{blue}\textbf{Group A2}} node[above=10pt] {};

  \draw[ultra thick] (-2.7,-1) node[below=3pt] {\color{red}\textbf{Group B1}} node[above=10pt] {};

  \draw[draw=black, fill=gray!20] (-3.2,0)--(-3.7,0.5)--(-1.8,0.5)--(-1.3,0)--(-1.8,-0.5)--(-3.7,-0.5)--(-3.2,0) node[xshift=5.3ex, align=center, text width=4.8em]{\textbf{Pre 1, \\Pre 2, OW}};

   \draw[draw=black, fill=gray!20] (-1.2,0)--(-1.7,0.5)--(0.2,0.5)--(0.7,0)--(0.2,-0.5)--(-1.7,-0.5)--(-1.2,0) node[xshift=5.3ex, align=center, text width=4.5em]{\textbf{LW 1--9}};

   \draw[draw=black, fill=gray!20] (0.8,0)--(0.3,0.5)--(2.2,0.5)--(2.7,0)--(2.2,-0.5)--(0.3,-0.5)--(0.8,0) node[xshift=5.3ex, align=center, text width=4.5em]{\textbf{X-mas}};
    
 \draw[draw=black, fill=gray!20] (2.8,0)--(2.3,0.5)--(4.2,0.5)--(4.7,0)--(4.2,-0.5)--(2.3,-0.5)--(2.8,0) node[xshift=5.5ex, align=center, text width=4.5em]{\textbf{LW 10--14}};

 \draw[draw=black, fill=gray!20] (4.8,0)--(4.3,0.5)--(6.2,0.5)--(6.7,0)--(6.2,-0.5)--(4.3,-0.5)--(4.8,0) node[xshift=5.3ex, align=center, text width=4.5em]{\textbf{EW 1 $\&$ 2}};

  \draw[draw=black, fill=gray!20] (6.8,0)--(6.3,0.5)--(8.2,0.5)--(8.7,0)--(8.2,-0.5)--(6.3,-0.5)--(6.8,0) node[xshift=5.3ex, align=center, text width=4.5em]{\textbf{Free 1--3}};

 \draw[draw=blue, fill=gray!60] (8.8,0)--(8.3,0.5)--(10.2,0.5)--(10.7,0)--(10.2,-0.5)--(8.3,-0.5)--(8.8,0) node[xshift=5.3ex, align=center, text width=4.5em]{\textbf{LW 1--14}};

 \draw[draw=blue, fill=gray!60] (10.8,0)--(10.3,0.5)--(12.2,0.5)--(12.7,0)--(12.2,-0.5)--(10.3,-0.5)--(10.8,0) node[xshift=5.3ex, align=center, text width=4.5em]{\textbf{EW 1}};

 \draw[draw=blue, fill=gray!60] (12.8,0)--(12.3,0.5)--(14.2,0.5)--(14.7,0)--(14.2,-0.5)--(12.3,-0.5)--(12.8,0) node[xshift=5.3ex, align=center, text width=4.5em]{\textbf{Free 1--6}};

 \draw[draw=red, fill=gray!20] (-3.2,-2.3)--(-3.7,-1.8)--(-1.8,-1.8)--(-1.3,-2.3)--(-1.8,-2.8)--(-3.7,-2.8)--(-3.2,-2.3) node[xshift=5.3ex, align=center, text width=4.8em]{\textbf{Pre 1, \\Pre 2, OW}};

   \draw[draw=red, fill=gray!20] (-1.2,-2.3)--(-1.7,-1.8)--(0.2,-1.8)--(0.7,-2.3)--(0.2,-2.8)--(-1.7,-2.8)--(-1.2,-2.3) node[xshift=5.3ex, align=center, text width=4.5em]{\textbf{LW 1--9}};

   \draw[draw=red, fill=gray!20] (0.8,-2.3)--(0.3,-1.8)--(2.2,-1.8)--(2.7,-2.3)--(2.2,-2.8)--(0.3,-2.8)--(0.8,-2.3) node[xshift=5.3ex, align=center, text width=4.5em]{\textbf{X-mas}};
    
 \draw[draw=red, fill=gray!20] (2.8,-2.3)--(2.3,-1.8)--(4.2,-1.8)--(4.7,-2.3)--(4.2,-2.8)--(2.3,-2.8)--(2.8,-2.3) node[xshift=5.3ex, align=center, text width=4.5em]{\textbf{LW 10--14}};

 \draw[draw=red, fill=gray!20] (4.8,-2.3)--(4.3,-1.8)--(6.2,-1.8)--(6.7,-2.3)--(6.2,-2.8)--(4.3,-2.8)--(4.8,-2.3) node[xshift=5.3ex, align=center, text width=4.5em]{\textbf{EW 1 $\&$ 2}};

  \draw[draw=red, fill=gray!20] (6.8,-2.3)--(6.3,-1.8)--(8.2,-1.8)--(8.7,-2.3)--(8.2,-2.8)--(6.3,-2.8)--(6.8,-2.3) node[xshift=5.3ex, align=center, text width=4.5em]{\textbf{Free 1--3}};

\end{tikzpicture}\vspace{-0.2cm}
    \caption{Overview of the data collection procedure for groups A1, A2, and B1 in the 1st semester (light gray) and 2nd semester (dark gray). (Pre = precourse, OW = orientation week, LW = lecture week, EW = exam week, Free = lecture-free week, and X-mas = Christmas break)}
    \label{fig:studydesign}
\end{figure*}
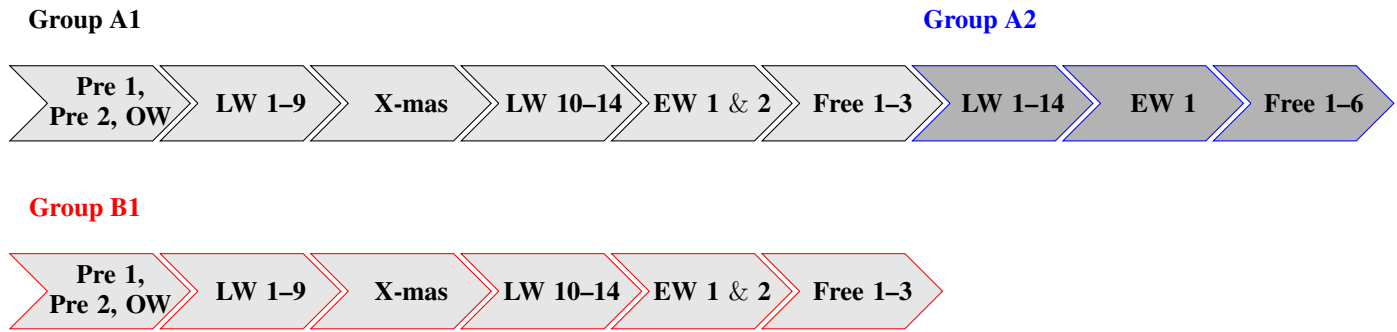

\begin{table*}[htb]
\caption{Sample overview. The total number of participants is the aggregation of all the different codes registered at any measuring point. Codes provided with slight variations were matched and considered as a single code. Demographics are presented where available. Since some students did not state their high school graduation grade, the $N$ is lower there. That grade is on a scale of 1.0 (best) to 4.0 (lowest).}
\begin{ruledtabular}
\footnotesize
\begin{tabular}{p{.29\textwidth}p{.22\textwidth}p{.22\textwidth}p{.22\textwidth}}
&Group A1 (winter semester 21/22)&Group A2 (summer semester 22)&Group B1 (winter semester 22/23)\\\hline
\textbf{Investigated group}&Cohort A, 1st semester&Cohort A, 2nd semester&Cohort B, 1st semester\\
Investigated course&Mathematical Methods in Physics&Experimental Physics II&Experimental Physics I\\\hline
\textbf{Total number of registered, matched codes}&145&116&164\\\hline
\textbf{Demographics available}&107&88&135\\
of which are in 1st resp. 2nd semester&97&64&125\\
of which are physics major&83&66&99\\
of which are teacher training students&18&15&25\\
of which are female&25&26&28\\
of which are of divers/unstated gender&3&1&7\\\hline
High school graduation grade &$M=1.7$, $SD=0.6$, $N=100$&$M=1.7$, $SD=0.6$, $N=83$&$M=1.8$, $SD=0.7$, $N=129$\\
\end{tabular}
\end{ruledtabular}
\label{Participants}
\end{table*}

In a panel study, stress perceptions among first-year physics students at the University of Göttingen were studied. The study focuses on the bachelor's program for physics majors and teacher training students. Students usually enroll in the three-year full-time program a few months or years after graduating from high school (traditional age students).

Figure~\ref{fig:StudyStructure} shows the standard study program for physics majors in their first year of study, i.e., in their first and second semesters. During this time, they take Experimental Physics~I and II and Mathematics for Physicists~I and II (or Integral and Differential Calculus~I and II). Additionally, they take Mathematical Methods in Physics in their first semester and Analytical Mechanics in their second semester. In each course, students attend one to two traditional large lectures (100 to 150 students) and a small group tutorial (15 to 25 students) each week. In all of these courses, except Mathematical Methods in Physics, students are required to complete and submit weekly exercise sheets, which are then graded by student assistants. At least half of all possible points are required throughout the semester as a prerequisite for the final written exam of two to three hours. Physics exams consist mostly of standard problem sets. They are high-stakes exams in which students must score a minimum number of points (usually between 45\% to 50\% of all possible points) to pass the course. Otherwise, they must retake the exam, whereby the number of retakes is limited, and multiple failures can lead to dismissal without graduation. If students pass the exam, they will receive an exam grade, which is the single, final grade for the entire course.

Additionally, students take an ungraded lab course both in their first and second semesters, in which they have to conduct several physics experiments and submit five written lab reports. A programming course is taken in the lecture-free time between the first and second semester. Before the first semester, students have the opportunity to attend a pre-course and an orientation week for onboarding, which most students take advantage of. The structure of the physics teacher training program is slightly different, as they have a second subject of study (e.g., mathematics or physical education).

The survey was conducted in three consecutive semesters as part of the undergraduate courses Experimental Physics~I and II and Mathematical Methods in Physics. In winter semester~21/22 (10/21-03/22) and summer semester~22 (04/22-09/22), data were collected from cohort~A of students in their first and second semesters, hereafter referred to as groups~A1 and A2. In winter semester~22/23 (10/22-03/23), cohort~B of students in their first semester was studied, hereafter referred to as group~B1. Examining these three groups allows us to compare how stress perceptions differ between first and second semesters and between two different cohorts.

As far as possible, data were collected via an online survey tool during the lectures within a short break, so the students could participate in the survey immediately and repeatedly. This mode of data collection was possible since one of the co-authors (PK) was the responsible lecturer in all investigated courses. Depending on the response rate during the lectures and, in particular, beyond the lecture times, the online survey link was sent to the students via e-mail. For group~A1 during lecture times, additionally, a paper version was used. However, for economic reasons and to enable data collection from students not regularly present on campus (e.g., during lecture-free periods or due to the COVID-19 pandemic), we subsequently transitioned to online surveys. This approach was adopted for all measurements with groups~A2 and B1.

As shown in Figure~\ref{fig:studydesign}, in the winter semesters (groups~A1 and B1), data were collected weekly in the two pre-course weeks and the orientation week, in each of the fourteen lecture weeks, during the Christmas break, and partly bi-weekly in the exam weeks and the lecture-free weeks thereafter. In summer semester 22 (group~A2), data were collected during the fourteen lecture weeks, once during the exam period, and partly bi-weekly in the lecture-free weeks thereafter; there was no pre-course or break during lecture time in that semester. This data collection scheme resulted in a total of sixty-seven measuring points, twenty-three for group ~A1, twenty-one for group~A2, and twenty-three for group~B1.

Pseudonymous codes were used to link responses from different measuring points. Potentially misspelled codes were matched if three out of four code elements were the same and if there was no obvious reason why these codes could not belong to the same person. Table~\ref{Participants} shows the number of matched codes, i.e., the number of participating students in the three groups and available demographic data. One hundred sixteen to 164 codes were identified per semester, and demographic data were given for 88 to 134 codes.

The majority of the participants in groups~A1 and B1 were actually in their first semester (92\%), and in group~A2 in their second semester (73\%). On average, 75\% of all participants were physics majors, 18\% were in physics teacher training, and 7\% were other types of students. Twenty-four percent of the students were female. The average high school graduation grade across all three groups is $M=1.7$, $SD=0.6$, $N=312$, where 1.0 would be the best and 4.0 the lowest possible grade. These students have very good high school graduation grades, as the average high school graduation grade in Lower Saxony ranged between 2.4 and 2.7 The high school graduation grade is, on average, across all three groups, $M=1.7$, $SD=0.6$, $N=312$ where 1.0 would be the best and 4.0 the lowest possible grade. These students have very good high school graduation grades as the recent average high school graduation grade in Lower Saxony was 2.43 \cite{Kultusministerkonferenz.2023}.

Overall, most participants fall into our primary target group of physics bachelor's degree students and teacher training students in their first or second semester. Therefore, we subject the entire data set to further analysis. The number of participants for each measuring point is shown in Figure~\ref{fig:Trajectory}. The full quantitative data set is available as supplementary material.

\subsection{Instrument}

\subsubsection{Overview, structure, and items}

The main instrument (cf. Table~\ref{tab:instrument} in the appendix) is the German version of the PSQ \cite{Fliege.2001}. All items can be interpreted in the context of studying physics, e.g., "\textit{You feel that too many demands are being made on you.}" (worries) or "\textit{You feel under pressure from deadlines.}" (tension). To provide a reference point, we instructed our students to relate the items to their current situation of studying physics. Thus, although we always surveyed students as part of a particular course throughout the semester, students were instructed to answer the questionnaire based on their overall impression of their studies in the previous week.

As in Refs.~\cite{Obermeier.2022,Handel.2020}, we used a six-point rating scale instead of the original four-point rating scale to improve the resolution for repeated measurements. In the online survey, items were presented in a randomized order. In addition to this slightly adapted PSQ, students were asked to assess the workload spent on their studies in the past week and to indicate up to three stressors that were most relevant to their stress perception in that week in open text fields.

\subsubsection{Instrument characterization} \label{instrumentevaluation}

\begin{figure}
    \centering
    \includegraphics[width=\columnwidth]{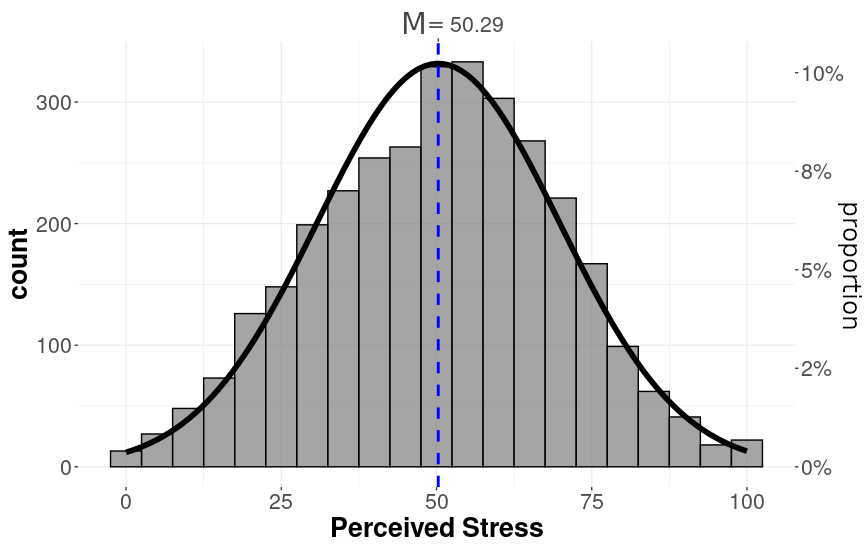}\vspace{-0.3cm}
    \caption{Distribution of the total perceived stress scores, based on the full data set ($N=3,241$) and with a normal distribution applied ($M=50.3,~SD=19.5$); the data set includes multiple responses from the same student at different measuring times.}
    \label{fig:histogram}
\end{figure}

Figure~\ref{fig:histogram} shows the distribution of total perceived stress scores across all participants and measuring points ($N=3,241$ observations). It reveals that students reported perceived stress across the entire scale from 0 to 100, with an average of $M=50.3$, $SE=0.34$, $SD=19.5$. Visual inspection of the empirical and theoretical cumulative distribution function (cf. Figure~\ref{fig:normaldistribution} in the appendix) and the D'Agostino test for skewness ($skew=-0.076,~z= -1.76,~p=.078$) indicate that this distribution can be considered normal.

\begin{figure*}
    \centering
    \includegraphics[width=.9\linewidth]{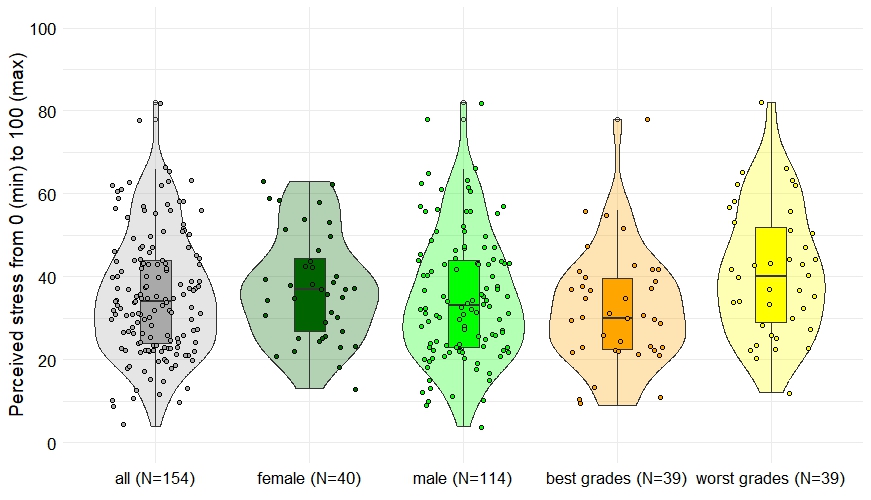}\vspace{-0.3cm}
    \caption{Violin plots for those $N=154$ students of the focus sample, i.e., Pre1 of groups~A1 and B1, for whom the high school graduation grade and information about whether they were female or male were provided. The plots visualize the scatter of the PSQ scores at the first measuring point for all students in that group and split by gender and the 25\% best and lowest high school graduation grades.}
    \label{fig:Violins}
\end{figure*}

To assess the internal consistency, Cronbach's alpha was calculated for a focus sample, Pre1 of groups~A1 and B1 ($N=196$ students), because these two measuring points are considered to be the least influenced by the study program, as they were only a few days after the students entered the university and the actual lecture time had not yet started. It is $\alpha=0.89$ which is sufficiently satisfactory and in line with the reference values by Ref.~\cite{Fliege.2001}, who have $\alpha=0.83$ for $N=246$ medical students at the beginning of their semester and $\alpha=0.85$ for their total sample group ($N=650$).

Based on the defined focus sample, we further analyze the scatter of the PSQ scores within a measuring point and the dependence on the two available predictors of interest, gender and high school graduation grade. Figure~\ref{fig:Violins} shows the scatter for the 154 students from the focus sample for whom we had both high school graduation grade and the information that they were either male or female. The PSQ scores range from 4 to 82, which is almost the entire scale from 0 to 100.

A t-test shows no significant differences ($t(152)=0.85$, $p=0.40$) between the PSQ scores of male students ($M=35.1$, $SE=1.4$, $N=114$) and female students ($M=37.4$, $SD=2.0$, $N=40$). Regarding high school graduation grades, we compare the quartile of $N=39$ students with the best ($M=1.0, SD=0.1$) and the lowest ($M=2.5$, $SD=0.3$) high school graduation grades. Between these two quartiles, the PSQ scores are significantly higher for students with the lowest high school graduation grades ($t(76)=2.72$, $p=0.008$) with an effect size of $d=0.62$, CI 95\% $[0.16, 1.07]$. Therefore, we do not differentiate by gender, but will briefly look at the relevance of high school graduation grades for stress perceptions during data analysis.

\subsubsection{Data interpretation}\label{datainterpretation}

A common approach to interpreting PSQ data, as used by Refs.~\cite{Bergdahl.2002,Kocalevent.2011,Heinen.2017}, is based on the mean and standard deviation of the usually single measuring point. A total score $\leq M+1~SD$ is interpreted as a \textit{mean} stress level, a score $> M+1~SD$ as a \textit{slightly increased} stress level, and $>M+2~SD$ as a \textit{high} stress level. To allow for a similar interpretation of the Göttingen data while accounting for repeated measurements, we derive these intervals from the focus sample with a total stress score of $M=35.7,\, SD=14.7$. Consequently, we interpret total stress scores between $[0,50]$ as \textit{mean} stress levels, between $(50,65]$ as \textit{slightly increased} stress levels, and between $(65,100]$ as \textit{high} stress levels.

\subsection{Method of data analysis}

\subsubsection{Quantitative data - stress perception \& workload (RQ 1)}\label{modificquandata}

Since not every student responded to the survey at every measuring point, the PSQ and workload data are further processed for statistical analysis. For this purpose, we focus on the pre-course, orientation week, and lecture weeks, since participation rates are significantly lower during the Christmas break, exam weeks, and lecture-free weeks (cf. Figure~\ref{fig:Trajectory}). This selection is also supported by the fact that the lecture period is the most relevant phase during the semester and that the requirements and circumstances are more similar among students during the lecture time than the lecture-free period.

To further deal with missing responses during the selected measuring points, we applied data smoothing by combining three consecutive weeks, resulting in five corresponding measuring periods throughout the semester: Period~1 is LW1-LW3, period~2 is LW4-LW6, period~3 is LW7-LW9, period~4 is LW10-12, and period~5 is LW13-LW14. For the winter semesters, an additional period~0 is defined, which represents the pre-course and orientation before the actual semester starts, i.e., the measuring points Pre1, Pre2, and OW. This results in a total of six measuring points in the winter semesters and five measuring points in the summer semester. The periods are shaded gray in Figures~\ref{fig:Trajectory} and \ref{fig:Workload}. The period number represents the same logical period within each semester. 

Since each period consists of three measuring points, every student could participate up to three times during each period. We average these up to three available total stress scores per participant to derive a stress score for each measuring period and student. If students did not participate in a single measuring period, the average of the corresponding measuring period of all other participants is imputed for the missing value. Students who did not participate in two or more measuring periods at least once were excluded from further analysis.

By this data processing, we obtain a data set for group~A1 consisting of fifty-two students for whom a stress score could be derived for all six measuring periods in the pre-lecture and lecture periods of the winter semester. For group~B1, the analogous data set consists of seventy-six students. For group~A2, the summer semester group consisting of only five measuring periods, the obtained data set consists of fifty-four students. Matching groups~A1 and A2 from the same cohort of students leaves a data set of thirty-nine students with stress scores for the six measuring periods in the winter semester and the five measuring periods in the summer semester.

The same procedure for smoothing and cleaning the data was applied analogously to the workload data. For each group, the final number of students after data processing is equal to the values reported in the previous paragraph.

The processed data sets were statistically analyzed to compare stress perceptions and self-reported workloads between the three groups and to examine progression over the course of the semester. A mixed ANOVA was used to compare groups~A1 and B1 (cohort~A vs. cohort B) with group as the between-subject factor and six measurement periods as the within-subject factor. Repeated measures ANOVAs were used to compare groups~A1 and A2 (first vs. second semester) with eleven measurement periods for the matched sample of thirty-nine students and to analyze changes between the measurement periods in the first and second semesters. According to Shapiro-Wilk tests for each group of interest, the data are normally distributed in most periods, but extreme outliers were found in several periods for the workload data, so a 90\% winsorization was applied to the workload data. For the mixed ANOVA, Levene's and Box's tests showed that the homogeneity of error variances and covariances was satisfied, except for the homogeneity of covariances for the total stress score ($p=0.011$). For the ANOVAs, the Greenhouse-Geisser adjustment was applied due to the lack of sphericity shown by Mauchly tests. A Bonferroni-adjusted post hoc analysis was performed for significant time trends.

\subsubsection{Qualitative data - stressors (RQ2)}

\begin{table*}[htb]
\caption{Overview of the category system for stressors listed by students in open text fields. It consists of three dimensions (University, Global, and Private) with categories each plus a category Miscellaneous. The full category system, including subcategories, more detailed descriptions, and anchor examples, is available as supplementary material.}
\begin{ruledtabular}
\footnotesize
\begin{tabular}{p{.025\textwidth}p{.205\textwidth}p{.73\textwidth}}
\multicolumn{2}{l}{Category}&Short description\\\hline
\multicolumn{2}{l}{\textbf{University}}&\\
U1&Study conditions&General (study) conditions at the university, in particular university guidelines, structures, (information) services as well as the human and material resources of the university\\
U2&Transition from school to university or start of semester&Transition from school to university or from the semester break to the new semester, e.g., lack of prior knowledge, catching up on school material or material from the previous semester, finding study groups\\
U3&Study organization&Individual organization of one's own studies, i.e., planning and structuring a semester or the entire course of study\\
U4&Individual daily study routines/time management&Daily study routine, especially individual weekly planning and time management\\
U5&Study-related self-regulation&Fears, insecurities, or pressures directly related to one's studies, including motivational factors and strategies\\
U6&Lab reports&Writing the lab reports/protocols as part of the physics lab course: requirements, number, scope, or deadlines, or unspecified\\
U7&Preparing and following up on lectures&Any work in addition to exercise sheets, etc. that serves to improve understanding of lecture content and fill in gaps in knowledge\\
U8&Lecture content&Subject matter in the lectures: novelty, difficulty, quantity, pace, or unspecific\\
U9&Exercise sheets&Weekly exercise sheets (problem sets): difficulty, quantity, or unspecified\\
U10&Exams and exam preparation&Exam admission \& exam preparation, exam phase in particular, or fear/anxiety of exams\\
U11&Unspecific mention of courses and subjects&Unspecified stress caused by their subject or courses attended, divided into math, physics, and lab courses, pre-course, and other subjects/courses\\
U12&Project group work&Stressors related to undergraduate research projects newly introduced only for group B1\\
\multicolumn{2}{l}{\textbf{Global}}&\\
G1&Financing of studies&Burden related to the own financing, including employment and general financial worries\\
G2&Future prospects&Study-unrelated (unspecified) fears, worries, and uncertainties about prospects for the present and future\\
G3&Covid-19 pandemic&Burden directly related to the Covid-19 pandemic\\
G4&Work-life balance&Burden on one's personal work-life balance, referring to private opportunities for relaxation or, conversely, when private obligations interfere with studies (in terms of time)\\
\multicolumn{2}{l}{\textbf{Private}}&\\
P1&Everyday demands&Burden caused by the demands of everyday life, including factors and especially activities related to (new) household management and living arrangements as well as extracurricular commitments, activities, and hobbies\\
P2&Private social environment&Social contacts in their personal environment or the establishment of a new social environment, including family, friends, and private contacts\\
P3&Illness&Temporary or long-term, physical or psychological illness, including unspecified descriptions of poor health states\\
\multicolumn{2}{l}{\textbf{Miscellaneous}}&Stressor that does not fall into any other category or cannot be meaningfully assigned to any of them\\
\end{tabular}
\end{ruledtabular}
\label{Categories}
\end{table*}

Across all measuring points, students reported 5,823 stressors, 2,216 by group~A1, 1,248 by group~A2, and 2,359 by group~B1. The responses were subjected to structuring qualitative content analysis, resulting in a category system summarized in Table~\ref{Categories}; the full category system is available as supplemental material. Analysis of the codings reveals the most salient stressors in different phases of the semester.

The category system consists of three dimensions, each of which is subdivided into categories and, in some cases, subcategories. These dimensions describe whether the stressors are primarily related to the \textit{university} and the students' studies (U), primarily \textit{private} (P), or \textit{globally} affect both the university and the private sphere (G). The \textit{university}-related dimension is further divided into eleven categories. Some of them describe typical activities of studying physics, i.e., writing \textit{lab reports} (U6), \textit{preparing and following up the lectures} (U7), understanding \textit{lecture content} (U8), solving regular \textit{exercise sheets} (U9), or \textit{preparing for and writing exams} (U10). Other categories contain the \textit{Study conditions} (U1), for which the university is responsible, the challenges related to the \textit{transition from school to university in the first semester or the start of the new semester} (U2), the general \textit{organization of one's own study program} (U3), the \textit{individual daily study routines} closely linked to the \textit{time management} (U4), and the \textit{study-related self-regulation} (U5) regarding motivation, emotions, thoughts, doubts, etc. Moreover, there is a category designated for all \textit{unspecific mentions of courses and subjects} (U11), i.e., for responses in which the exact underlying stressors are not clarified. There is also a category for \textit{project group work} (U12), which was introduced in the winter semester 22/23. The \textit{private} dimension is divided into three categories dealing with \textit{everyday demands} (P1) such as household, private activities, etc., the \textit{private social environment} (P2) and relationships, and \textit{illness} (P3). The \textit{global} dimension is divided into four categories related to \textit{financing of one's studies} (G1), fears/uncertainties about general \textit{future prospects} (G2), the \textit{COVID-19 pandemic} (G3), and \textit{work-life balance} (G4) including relaxation. An additional miscellaneous dimension/category is used to categorize all other responses can be categorized that cannot be clearly assigned to the previous categories due to their lack of specificity or rarity.

To assess the reliability of the category system, an interrating was conducted involving the first author (rater 1) and three 2$^{\text{nd}}$ to 3$^{\text{rd}}$-year teacher training students (raters~2 to 4) to integrate the student's perspective into the data analysis. Around 31\% of the responses were subjected to interrating, with care taken to ensure a balanced distribution of data across different measuring points in the semester and across the three groups. Each rater independently coded subsamples of responses. Differences in ratings were discussed among the three raters each involved. Table~\ref{interrating} presents Cohen's kappa for two and Fleiss' kappa for three raters before ($\kappa_{pre}$) and after ($\kappa_{post}$) discussion of codings at the subcategory level of the category system. All $\kappa_{pre}$-values, except Fleiss' kappa for group B1, exceed the 0.61 threshold indicating \textit{substantial} agreement; all $\kappa_{post}$-values surpass the 0.80 threshold and can therefore be interpreted as \textit{almost perfect} \cite{Landis.1977}. The discussions led to various minor refinements of the category system including clarifications and distinctions between categories implemented in the current category system and codings.

\begin{table}[htb]
\caption{Cohen's $\kappa$ for two raters and Fleiss's $\kappa$ for three raters, measure the agreement on approximately 31\% of the mentioned stressors coded using the category system before ($\kappa_{pre}$) and after ($\kappa_{post}$) discussions between the raters involved. Rater~1 was the first author, raters~2 to 4 were bachelor teacher training students.}
\begin{ruledtabular}
\footnotesize
\begin{tabular}{p{.28\columnwidth}p{.105\columnwidth}p{.105\columnwidth}p{.105\columnwidth}p{.105\columnwidth}p{.105\columnwidth}p{.105\columnwidth}}
Comparison&\multicolumn{2}{l}{Group A1}&\multicolumn{2}{l}{Group A2}&\multicolumn{2}{l}{Group B1}\\
&$\kappa_{pre}$&$\kappa_{post}$&$\kappa_{pre}$&$\kappa_{post}$&$\kappa_{pre}$&$\kappa_{post}$\\\hline
Rater 1 vs 2&0.80&0.92&&&\\
Rater 1 vs 3&0.78&0.92&0.76&0.97&&\\
Rater 2 vs 3&0.76&0.91&&&0.71&0.93\\
Rater 1 vs 2 vs 3&0.71&0.88&&&&\\
Rater 1 vs 4&&&0.79&0.95&&\\
Rater 3 vs 4&&&0.74&0.96&0.70&0.95\\
Rater 1 vs 3 vs 4&&&0.68&0.95&&\\
Rater 2 vs 4&&&&&0.72&0.94\\
Rater 2 vs 3 vs 4&&&&&0.59&0.90\\
\end{tabular}
\end{ruledtabular}
\label{interrating}
\end{table}

\section{Results}

\subsection{Trajectory of perceived stress (RQ1a)}

Figure~\ref{fig:Trajectory} shows the trajectory of perceived stress, i.e., the average stress level of all participating students for each measuring point on the scale from 0 (min.) to 100 (max.) over the semester for the groups~A1, A2, and B1. The three trajectories exhibit a similar pattern. In the winter semesters (groups~A1 and B1), the students began with a perceived stress level of around thirty-five in the pre-course. In the first four to five lecture weeks, the stress level increased in all groups to a level of around fifty-five, a level sustained with minor fluctuations until the exam weeks. After the exams, the stress level decreased even below the starting level, down to around twenty to thirty. In group~A1, the stress perception also diminished during the Christmas break. In group~A2, measuring points in the later lecture-free time indicate a minor increase in the stress level.

The following statistical analysis is based on the processed data set as described in Sec.~\ref{modificquandata}. The corresponding more abstract trajectory of perceived stress is shown in Figure~\ref{fig:ModifiedTrajectory} in the Appendix. In addition, in Figure~\ref{fig:ModifiedTrajectorySplit}, we visualize the same processed trajectory of perceived stress for a subset of students from groups~A1 and B1 for whom the high school graduation grade was available. Data are presented for the quartiles of students with the best ($M=1.03$, $SD=0.05$) and the lowest ($M=2.17$, $SD=0.31$) high school graduation grades. Visual inspection already reveals that there is no significant difference in perceived stress scores during the entire lecture time (periods~1 to 5), except for period~0,  between students with the best and lowest high school graduation rates. Therefore, the following statistical analysis is performed without differentiating by high school graduation grades.

\subsubsection{Differences between the three groups}\label{StressDiffgroups}

For groups~A1 and B1, both cohorts in their first semester, a mixed ANOVA supports the impression that the course of stress perception is similar for both groups. There are neither interaction effects between the measuring period and group ($F(4.2, 526.1) = 1.3, \, p=0.26$, $\eta^2_P= 0.01$) nor differences between the groups ($F(1,126) = 0.5, \, p=0.48$, $\eta^2_P= 0.003$). Thus, we merge the data sets of groups~A1 and B1 when analyzing the temporal progression of perceived stress throughout the first semester in the next section.

For the matched sample of groups~A1 and A2, a repeated measures ANOVA shows a significant temporal evolution of perceived stress within the two semesters ($F(5.7, 215.6) = 40.0, \, p<0.001$, $\eta^2_P= 0.51$). Post-hoc analysis, in which we first focus on a comparison of corresponding measurement periods, shows different levels of perceived stress between the semesters during period~1 but equal stress levels thereafter. The total stress score in period~1 of group~A2 is significantly lower than the score in period~1 of group~A1 ($p=0.003$) but still statistically equal to period~0 of group~A1 ($p=1.0$). For all further measuring periods, the level of perceived stress during corresponding periods is statistically equal ($p=1.0$ for periods~2-4; $p=0.145$ for period~5).

\subsubsection{Differences between different phases of the semester}\label{stressdiffperiods}

The level of perceived stress increased throughout the course of both the first and second semesters. Repeated measures ANOVAs show significant variation of the total stress score with similar effect sizes (first semester, i.e., groups~A1 and B1 combined: $F(4.2,529.1) = 137.1, p<0.001$, $\eta^2_P= 0.52$; second semester, i.e., group~A2: $F(2.8, 149.4) = 41.1, p<0.001$, $\eta^2_P= 0.44$). 
For the first semester, post-hoc analysis (cf. $p$-values in Table~\ref{tab:pvalues} in the appendix) shows that the perceived stress during periods~0 and 1, i.e., between pre-lecture time and the first lecture weeks, differs significantly from one another as well as from all further periods in the lecture time ($p<0.001$). Thereafter, no significant differences between the periods occur. Thus, in the first semester, the level of perceived stress increased during the beginning of the semester but reached a constant plateau after only three lecture weeks. Similarly, within the second semester, the stress level in period~1 significantly differs from all further measuring periods. Additionally, the stress score in period~5 was significantly higher than in periods~2 and 3 ($p<0.001$), i.e., unlike for the first semester, the stress perception continued to rise in the last two lecture weeks of the second semester.

\begin{figure*}
\centering
\begin{tikzpicture}
\begin{axis}[ width=\textwidth, height=9cm, 
grid=none, line width= 1pt,
  ymax=75,  
  xmin=0.5, xmax=26.5,
  ytick={0,10,20,30,40,50,60,70,80,90,100},
  ymin =0, 
  ylabel={Perceived stress from 0 (min) to 100 (max)},
  xtick = {1,2,3,4,5,6,7,8,9,10,11,12,13,14,15,16,17,18,19,20,21,22,23,24,25,26},
  xticklabel style={text width=1.2cm, anchor=east, rotate=90, },
  xticklabels={Pre 1, Pre 2, OW, LW 1, LW 2, LW 3, LW 4, LW 5, LW 6, LW 7, LW 8, LW 9, X-mas, LW 10, LW 11,  LW 12, LW 13, LW 14, EW 1, EW 2, Free 1, Free 2, Free 3, Free 4, Free 5, Free 6},
   xtick pos=left,
ytick pos=left,
extra x ticks={1,2,3,4,5,6,7,8,9,10,11,12,13,14,15,16,17,18,19,20,21,22,23,24,25,26},
extra x tick labels={93\newline\newline103,81\newline\newline48,76\newline\newline59,108\newline86\newline128,87\newline80\newline81,71\newline42\newline45,85\newline40\newline81,81\newline73\newline71,66\newline63\newline78,47\newline46\newline66,42\newline53\newline45,41\newline45\newline18,26\newline\newline27,39\newline37\newline86,40\newline46\newline59,47\newline3\newline79,39\newline34\newline67,44\newline41\newline18,23\newline21\newline23,21\newline\newline17,22\newline17\newline15,16\newline14\newline14,14\newline12\newline6,\newline10\newline,\newline12\newline,\newline23\newline},
extra x tick style={grid=none, tick style={draw=none}, tick label style={xshift=32pt, yshift=22pt, rotate=270}},
 legend columns=4, legend cell align = left,legend style = {draw = none},
 legend style = {at ={(0,1)}, anchor = south west},
 bar width = 5pt,
  ]
    \addlegendimage{empty legend}\addlegendentry{\scalebox{1}[1]{\ref{StressA1}} Group A1
    }
    \addlegendimage{empty legend}\addlegendentry{\scalebox{1}[1]{\ref{StressA2}} Group A2
    }
    \addlegendimage{empty legend}\addlegendentry{\scalebox{1}[1]{\ref{StressB1}} Group B1
    }

\draw[fill=gray,draw=gray!20, fill opacity=0.2] (0,0) rectangle (30,750);
\draw[fill=gray,draw=gray!20, fill opacity=0.1] (30,0) rectangle (60,750);
\draw[fill=gray,draw=gray!20, fill opacity=0.2] (60,0) rectangle (90,750);
\draw[fill=gray,draw=gray!20, fill opacity=0.1] (90,0) rectangle (120,750);
\draw[fill=gray,draw=gray!20, fill opacity=0.2] (130,0) rectangle (160,750);
\draw[fill=gray,draw=gray!20, fill opacity=0.1] (160,0) rectangle (180,750);

\node[draw] at (15,250) {Period 0};
\node[draw] at (45,250) {Period 1};
\node[draw] at (75,250) {Period 2};
\node[draw] at (105,250) {Period 3};
\node[draw] at (145,250) {Period 4};
\node[draw] at (175,250) {Period 5};
\node[] at (30,170) {Number of participants};

\addplot[scatter/classes={a={black}},
    scatter, visualization depends on=\thisrow{ey} \as \myshift,
    every node near coord/.append style = {shift={(axis direction
    cs:0,\myshift)}},
    scatter src=explicit symbolic,
    ]
    plot [mark=*,draw=black, thick, mark options = {solid}, error bars/.cd, y dir = both, y explicit,  error mark options={solid, black, rotate=90,mark size=2pt,}, error bar style ={solid}]
    table[meta=class, x=x, y=y, y error=ey]{
        x   y   ey    class label
        1   35.78324844  1.51950941 1
       2	34.32358674 1.699826361 2
       3	36.43767313 1.855920287 3
       4   44.2202729 1.5248668 4
       5	48.83721 1.560521994 5
       6 55.3195 1.815631102 6
        7  56.61733746 1.859610469 7
       8   57.80181936 1.832795856 8
        9   57.80701754 1.956134327 9
        10  62.54647256 2.753355315 10
        11 58.5 2.886198083 11
     12 61.36585366 2.879650237 12
     13 45.26923077 3.896569573 13
     14 57.87179487 3.150558395 14
     15 61.125 3.332279481 15
     16 61.94512878 2.737746875 16
     17 58.33333333 3.00029988 17
          18 58.22727273 2.857589508 18
           19 62.86956522 4.043881998 19
           20 53.66666667 5.57858718 20
        21 27.68181818 4.799164052 21
        22 20.75 4.675735236 22
        23 30 6.130450031 23
        }; \label{StressA1}
       
\addplot+[scatter/classes={a={red}},
    scatter, visualization depends on=\thisrow{ey} \as \myshift,
    every node near coord/.append style = {shift={(axis direction
    cs:0,\myshift)}},
    scatter src=explicit symbolic,
    ]
    plot [mark=*,draw=red, thick, mark options = {solid}, error bars/.cd, y dir = both, y explicit,  error mark options={solid, red, rotate=90,mark size=2pt,}, error bar style ={solid}]
    table[meta=class, x=x, y=y, y error=ey]{
        x   y   ey    class label
        4 40.12643678 2.209599999 4
        5 47.55 2.020926909 5
        6 48.52380952 2.991322738 6
        7 51.775 2.639319915 7
        8 53.19178082 1.857116135 8
        9 55.50793651 2.271269499 9
        10 55.7173913 2.647150148 10
        11 53.94339623 2.397448918 11
        12 51 2.321833132 12
        }; \label{StressA2}

        \addplot+[scatter/classes={a={red}},
    scatter, visualization depends on=\thisrow{ey} \as \myshift,
    every node near coord/.append style = {shift={(axis direction
    cs:0,\myshift)}},
    scatter src=explicit symbolic,
    ]
    plot [mark=*,draw=red, thick, mark options = {solid, fill=red}, error bars/.cd, y dir = both, y explicit,  error mark options={solid, red, rotate=90,mark size=2pt,}, error bar style ={solid}]
    table[meta=class, x=x, y=y, y error=ey]{
        x   y   ey    class label
        14 53.18918919 2.702417402 14
        15 58.23913043 2.416766071 15
        16 61.66666667 8.838049056 16
        17 61.02941176 2.889986799 17
        18 63.48780488 2.667355032 18
        19 50.19047619 4.319181496 19
        }; 

        \addplot+[scatter/classes={a={red}},
    scatter, visualization depends on=\thisrow{ey} \as \myshift,
    every node near coord/.append style = {shift={(axis direction
    cs:0,\myshift)}},
    scatter src=explicit symbolic,
    ]
    plot [mark=*,draw=red, thick, mark options = {solid, fill=red}, error bars/.cd, y dir = both, y explicit,  error mark options={solid, red, rotate=90,mark size=2pt,}, error bar style ={solid}]
    table[meta=class, x=x, y=y, y error=ey]{
        x   y   ey    class label
        21 30.29411765 5.186779508 21
        22 23.64285714 3.748416725 22
        23 34.58333333 6.035825285 23
        24 36.7 8.145960144 24
        25 32.41666667 5.710779724 25
        26 31.91304348 4.749001577 26
        }; 

\addplot[scatter/classes={a={blue}},
    scatter, visualization depends on=\thisrow{ey} \as \myshift,
    every node near coord/.append style = {shift={(axis direction
    cs:0,\myshift)}},
    scatter src=explicit symbolic,
    ]
    plot [mark=*,draw=blue, thick, mark options = {solid, fill=blue}, error bars/.cd, y dir = both, y explicit,  error mark options={solid, blue, rotate=90,mark size=2pt,}, error bar style ={solid}]
    table[meta=class, x=x, y=y, y error=ey]{
        x   y   ey    class label
        1 35.66990291 1.453013591 1
        2 38.52083333 2.159834341 2
        3 35.61016949 1.745666246 3
        4 45.515625 1.31475876 4
        5 50.09876543 1.88296497 5
        6 58.44444444 2.285782927 6
        7 54.56790123 1.694484359 7
        8 56.61971831 1.888583321 8
        9 56.67948718 1.975330167 9
        10 56.18181818 1.926373102 10
        11 59.08888889 2.577059287 11
        12 56.27777778 2.570325322 12
        13 55.14814815 4.986221386 13
        14 57.34883721 1.821273246 14
        15 60.54237288 1.971053238 15
        16 53.92405063 1.773783271 16
        17 55.53731343 1.959486586 17
        18 62.66666667 2.584519019 18
        19 63.60869565 2.789547919 19
        20 40.64705882 5.487245356 20
        21 26.6 4.787682609 21
        22 32.5 5.306796389 22
        23 20.16666667 6.477739661 23
        }; \label{StressB1}
\end{axis}
\end{tikzpicture}\vspace{-0.3cm}
\caption{Perceived stress (mean and standard error for each measuring point) on the artificial scale from 0 (min.) to 100 (max.) of groups~A1, A2, and B1, i.e., cohorts~A and B in their first or second semester, over the semester with the precourse (Pre), orientation week (OW), lecture weeks (LW), exam weeks (EW), lecture-free weeks (Free), and Christmas break (X-mas) in the winter semesters. The numbers below indicate the number of participants per measuring point and group (first row for group~A1, second row for group~A2, and third row for group~B1). The gray areas visualize which measuring points were combined into periods~0 to 5 for quantitative data analysis.}
\label{fig:Trajectory}
\end{figure*}
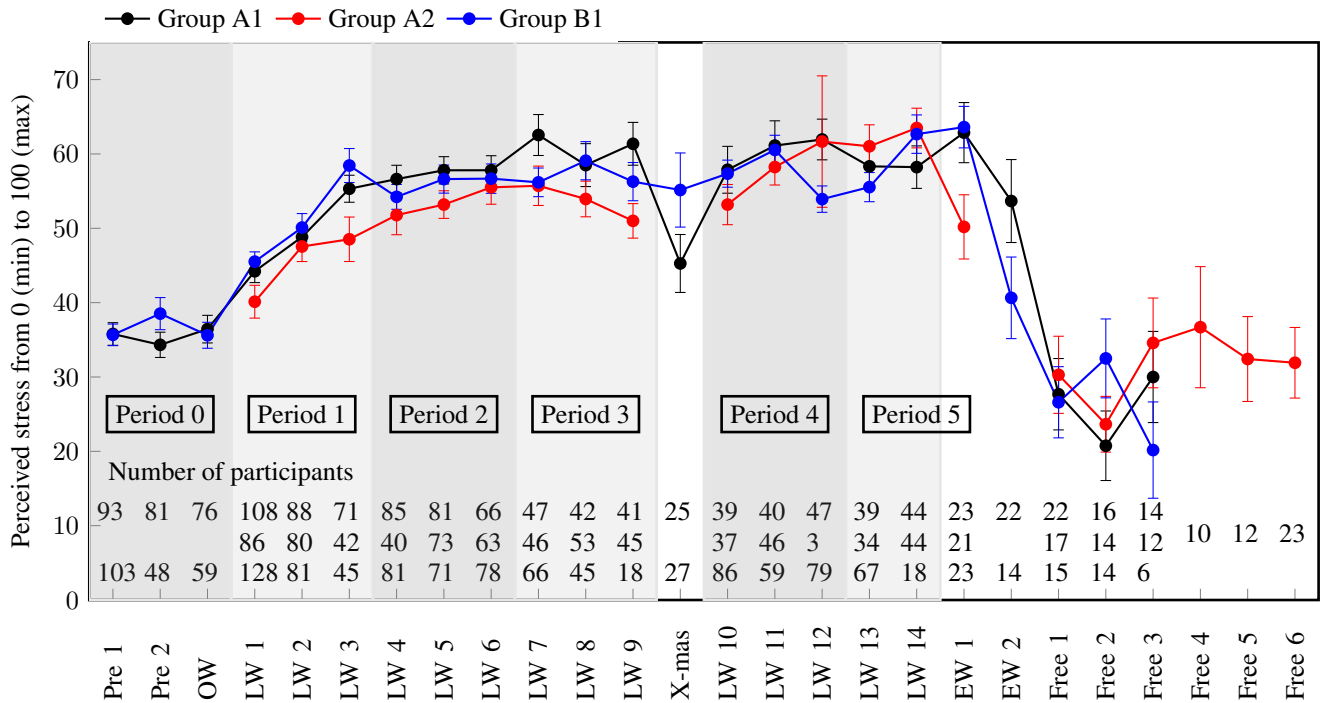

\subsection{Self-reported workload (RQ1b)}

\begin{figure*}[hbtp]
\centering  
\begin{tikzpicture}
\begin{axis}[ width=\textwidth, height=8cm, 
grid=none, line width= 1pt,
  ymax=65,  
  xmin=0.5, xmax=26.5,
  ytick={0,10,20,30,40,50,60,70,80,90,100},
  ymin =0, 
  ylabel={Self-estimated, weekly study-related workload (in h)},
  xtick = {1,2,3,4,5,6,7,8,9,10,11,12,13,14,15,16,17,18,19,20,21,22,23,24,25,26},
  xticklabel style={text width=1.2cm, anchor=east, rotate=90 },
  xticklabels={Pre 1, Pre 2, OW, LW 1, LW 2, LW 3, LW 4, LW 5, LW 6, LW 7, LW 8, LW 9, X-mas, LW 10, LW 11,  LW 12, LW 13, LW 14, EW 1, EW 2, Free 1, Free 2, Free 3, Free 4, Free 5, Free 6},
   xtick pos=left,
ytick pos=left,
 legend columns=4, legend cell align = left,legend style = {draw = none},
 legend style = {at ={(0,1)}, anchor = south west},
 bar width = 5pt,
  ] 
    \addlegendimage{empty legend}\addlegendentry{\scalebox{1}[1]{\ref{workloadA1}} Group A1
    }
    \addlegendimage{empty legend}\addlegendentry{\scalebox{1}[1]{\ref{workloadA2}} Group A2
    }
    \addlegendimage{empty legend}\addlegendentry{\scalebox{1}[1]{\ref{workloadB1}} Group B1
    }
\draw[fill=gray,draw=gray!20, fill opacity=0.2] (0,0) rectangle (30,750);
\draw[fill=gray,draw=gray!20, fill opacity=0.1] (30,0) rectangle (60,750);
\draw[fill=gray,draw=gray!20, fill opacity=0.2] (60,0) rectangle (90,750);
\draw[fill=gray,draw=gray!20, fill opacity=0.1] (90,0) rectangle (120,750);
\draw[fill=gray,draw=gray!20, fill opacity=0.2] (130,0) rectangle (160,750);
\draw[fill=gray,draw=gray!20, fill opacity=0.1] (160,0) rectangle (180,750);

\node[draw] at (15,45) {Period 0};
\node[draw] at (45,45) {Period 1};
\node[draw] at (75,45) {Period 2};
\node[draw] at (105,45) {Period 3};
\node[draw] at (145,45) {Period 4};
\node[draw] at (175,45) {Period 5};

\addplot[scatter/classes={a={black}},
    scatter, visualization depends on=\thisrow{ey} \as \myshift,
    every node near coord/.append style = {shift={(axis direction
    cs:0,\myshift)}},
    scatter src=explicit symbolic,
    ]
    plot [mark=*,draw=black, thick, mark options = {solid}, error bars/.cd, y dir = both, y explicit,  error mark options={solid, black, rotate=90,mark size=2pt,}, error bar style ={solid}]
    table[meta=class, x=x, y=y, y error=ey]{
        x   y   ey    class label
1	23.51612903	1.212286418 1
2	22	0.872161527 2
3	10.88461538	1.475149455 3
4	21.79807692	1.051435194 4
5	40.64457831	1.613886855 5
6	46.36764706	1.559775862 6
7	47.07037037	1.264602291 7
8	48.81351351	1.4286081 8
9	54.47142857	1.327880643 9
10	55.65319149	1.780877589 10
11	53.5175	1.913096332 11
12	55.41315789	1.844910229 12
13	22.92307692	2.234223321 13
14	42.3	2.643784246 14
15	53.91538462	1.887286197 15
16	53.34782609	1.769779779 16
17	52.8974359	1.89494493 17
18	53.20454545	1.789852636 18
19	51.52173913	2.819237671 19
20	50.9047619	3.740960382 20
21	18.38636364	3.607041547 21
22	13.625	3.804465578 22
23	22.14285714	6.009934214 23
        }; \label{workloadA1}
        
\addplot+[scatter/classes={a={red}},
    scatter, visualization depends on=\thisrow{ey} \as \myshift,
    every node near coord/.append style = {shift={(axis direction
    cs:0,\myshift)}},
    scatter src=explicit symbolic,
    ]
    plot [mark=*,draw=red, thick, mark options = {solid, fill=red}, error bars/.cd, y dir = both, y explicit,  error mark options={solid, red, rotate=90,mark size=2pt,}, error bar style ={solid}]
    table[meta=class, x=x, y=y, y error=ey]{
        x   y   ey    class label
4	9.23255814	0.997901275 4
5	38.34375	1.660924559 5
6	45	2.260091577 6
7	46.95	2.300209021 7
8	44.71232877	1.71765382 8
9	48.3015873	2.00672639 9
10	50.10869565	2.066579769 10
11	45.46226415	2.354852056 11
12	47.93333333	2.042577101 12
        }; \label{workloadA2}
        
\addplot+[scatter/classes={a={red}},
    scatter, visualization depends on=\thisrow{ey} \as \myshift,
    every node near coord/.append style = {shift={(axis direction
    cs:0,\myshift)}},
    scatter src=explicit symbolic,
    ]
    plot [mark=*,draw=red, thick, mark options = {solid, fill=red}, error bars/.cd, y dir = both, y explicit,  error mark options={solid, red, rotate=90,mark size=2pt,}, error bar style ={solid}]
    table[meta=class, x=x, y=y, y error=ey]{
        x   y   ey    class label
14	49.41891892	1.975973694 14
15	52.08695652	1.850140636 15
16	53.33333333	8.006941433 16
17	51.61764706	2.648113278 17
18	54.6097561	2.57731143 18
19	48.23809524	3.335169562 19
}; 

\addplot+[scatter/classes={a={red}},
    scatter, visualization depends on=\thisrow{ey} \as \myshift,
    every node near coord/.append style = {shift={(axis direction
    cs:0,\myshift)}},
    scatter src=explicit symbolic,
    ]
    plot [mark=*,draw=red, thick, mark options = {solid, fill=red}, error bars/.cd, y dir = both, y explicit,  error mark options={solid, red, rotate=90,mark size=2pt,}, error bar style ={solid}]
    table[meta=class, x=x, y=y, y error=ey]{
        x   y   ey    class label
21	19.29411765	4.535614546 21
22	14.42857143	4.047403571 22
23	25.84615385	3.133455031 23
24	20.5	6.4743082 24
25	18.75	5.971047571 25
26	12.34782609	3.667832984 26
}; 

\addplot[scatter/classes={a={blue}},
    scatter, visualization depends on=\thisrow{ey} \as \myshift,
    every node near coord/.append style = {shift={(axis direction
    cs:0,\myshift)}},
    scatter src=explicit symbolic,
    ]
    plot [mark=*,draw=blue, thick, mark options = {solid, fill=blue}, error bars/.cd, y dir = both, y explicit, error mark options={solid, blue, rotate=90,mark size=2pt,}, error bar style ={solid}]
    table[meta=class, x=x, y=y, y error=ey]{
        x   y   ey    class label
1	18.94174757	1.388441534 1
2	25.03125	1.409221644 2
3	10.175	1.239481745 3
4	21.6796875	1.188606357 4
5	36.30864198	1.574422826 5
6	45.88888889	1.868389957 6
7	48.05	1.610740176 7
8	46.26408451	1.750178026 8
9	51.44871795	1.740174886 9
10	52.86923077	1.66085041 10
11	55.36666667	2.972385061 11
12	54.15277778	4.758254733 12
13	25.96296296	2.702083189 13
14	41.65988372	1.781853082 14
15	57.05084746	1.57968067 15
16	51.54746835	1.537573856 16
17	52.49253731	1.589883169 17
18	50.05555556	3.289388983 18
19	50.75	3.973012964 19
20	43.47058824	5.955791575 20
21	12.31666667	2.762037148 21
22	22.35714286	4.348306659 22
23	3.75	2.379600807 23
        }; \label{workloadB1}
\end{axis}
\end{tikzpicture}\vspace{-0.3cm}
\caption{Workload (mean and standard error for each measuring time) in hours reported by groups~A1, A2, and B1, i.e., cohorts~A and B in their first or second semester, that they spent each week during the semester on their studies. The gray areas visualize which measuring points were combined into periods~0 to 5 for quantitative data analysis.}
\label{fig:Workload}
\end{figure*}
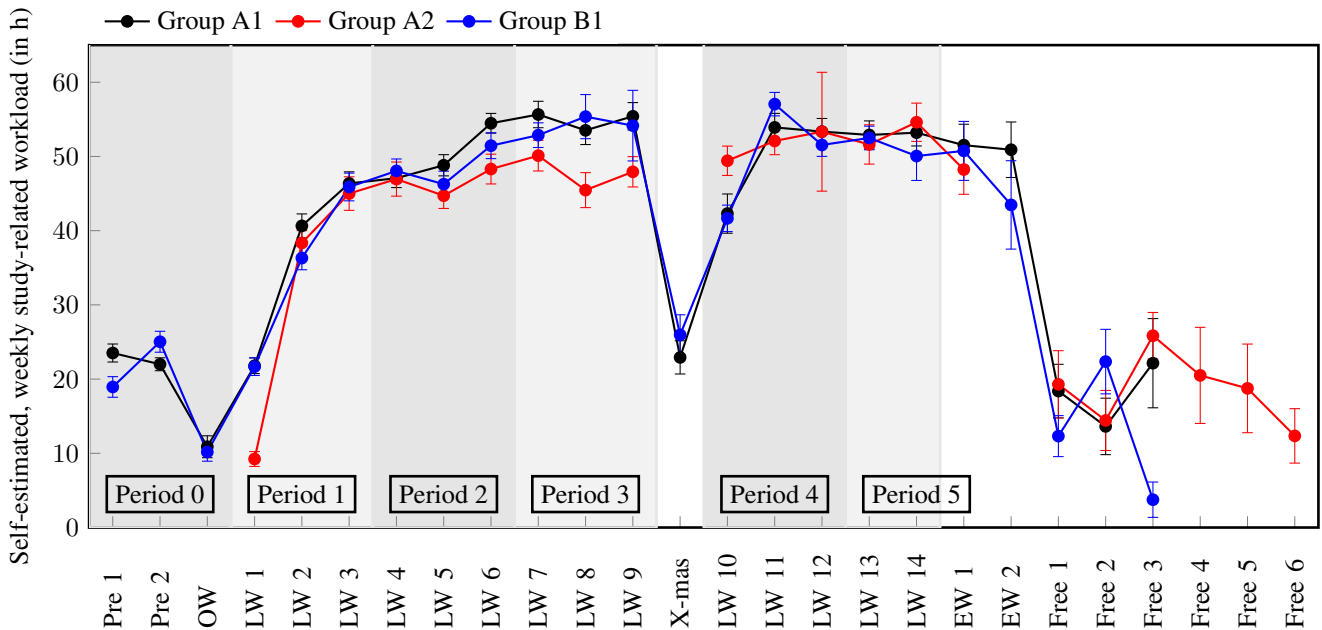

Figure~\ref{fig:Workload} shows the weekly study-related workload in hours according to students' self-estimation over the semester. The three groups reported similar workloads in corresponding weeks of the semester. During the pre-course (groups~A1 and B1), students reported that they spent about twenty hours on their studies. During the first four to six lecture weeks, this self-estimated workload increased strongly to a level of about fifty hours per week in all groups. This level continued until the end of the exam weeks, after which the reported workload dropped back to a level of around twenty hours or even less. During the Christmas break, students indicated a much lower workload of about twenty hours. The following statistical analysis is based on the processed data set as described in Sec.~\ref{modificquandata}; the corresponding more abstract trajectory of self-reported workload is shown in Figure~\ref{fig:Modifiedworkload} in the appendix.

\subsubsection{Differences between the three groups}

Comparing the workloads of groups~A1 and B1, a mixed ANOVA shows neither interaction effects of temporal evolution and group ($F(4.1, 518.1) = 2.2$, $p=0.06$, $\eta^2_P = 0.02$) nor group effects ($F(1, 126) = 2.8$, $p=0.098$, $\eta^2_P = 0.02$). Consequently, we combine the data from groups~A1 and B1 again when analyzing temporal effects in the next section.

For the matched sample of groups~A1 and A2, a repeated measures ANOVA shows a significant temporal effect of workload ($F(6.3, 241.1) = 108.5,$ $p<0.001$, $\eta^2_P = 0.74$). Post-hoc analysis reveals that period~1 of group~A2 is significantly different from all other measuring periods ($p<0.02$). In particular, this implies that the average workload was significantly lower in period~1 of group~A2 than in the corresponding period of group~A1, but significantly higher than in period~0 of group~A1. There are no significant differences between the reported workloads of other corresponding measuring periods within the course of the semester ($p>0.9$ in all pairwise comparisons).

\subsubsection{Differences between different phases of the semester}\label{workloaddiffperiods}

A repeated measures ANOVA of the reported workload data for the first semester (combined groups~A1 and B1) reveals significant changes in workload over the course of the lecture weeks ($F(4.1,521.3)=405.3$, $p<0.001$, $\eta^2_P = 0.76$). Post-hoc pairwise comparisons (cf. Table~\ref{tab:pvalues} in the appendix) show that pre-lecture period~0 and period~1 are significantly different from each other and from later periods ($p<0.001$). Moreover, periods~3 and 4 ($p=0.002$) and periods~4 and 5 ($p=0.002$) are significantly different from each other, as the mean reported workload in period~4 is below the reported values in the adjacent measuring periods. All other reported values are statistically comparable. For the second semester (group~A2), a repeated measures ANOVA reveals a significant change in the reported workload within the lecture time ($F(3.4,181.6) = 111.7$, $p<.001$, $\eta^2_P = 0.68$). Post-hoc analysis shows that the workload in period~1 is significantly different from all later measuring periods ($p<0.001$). Moreover, period~5 is significantly different from period~2 ($p=0.024$) and period~3 ($p=0.030$). The other pairwise comparisons are significant ($p>0.60$).

\subsubsection{Correlation between perceived stress and workload}\label{correlationstressworkload}

The repeated measures correlation \cite{Bakdash.2017} between the total stress score and the reported workload was calculated over all measuring points and all groups, i.e., $N=3,198$ observations. It is $r_{rm}(2843)=0.62$, CI $99\%\;[0.59, 0.65]$, $p<0.001$.

\subsection{Stressors contributing to stress perception (RQ2)}

\subsubsection{Total frequency of various stressors}

Figure~\ref{fig:stressors} presents the percentage of codings per category and group. In all three groups, the category \textit{exercise sheets} (U9) was mentioned most frequently with a share of 15\% to 21\%. The second and third most common stressors with a share of 9\% to 15\% were the categories \textit{exams and exam preparation} (U10) and \textit{unspecific mention of courses/subjects} (U11), i.e., a course or subject was mentioned without further information about the underlying stressors. Altogether, these three categories represent 35\% to 49\% of all responses in the three groups. All other categories were coded less frequently for up to 8\% of all responses per group. Each category was used in at least 1\%, and sixteen out of twenty categories were used in at least 3\% of the codings for a group.

\begin{figure}[htb]
\flushleft 
\begin{tikzpicture}
\begin{axis}[width=.73\columnwidth, height=18cm, xbar=0pt,
  ymax=4.1,  
  xmin=0, xmax=21,
  xtick={0,5,10,15,20},
  ymin =0.1,
  xlabel={Percentage (\%)},
  ytick = {0.2,0.4,.6,.8,1.,1.2,1.4,1.6,1.8,2.,2.2,2.4,2.6,2.8,3,3.2,3.4,3.6,3.8,4},
  yticklabel style={text width=.4\columnwidth,align=right, },
  yticklabels={Miscellaneous, P3 Illness, P2 Priv. social environment, P1 Everyday demands, G4 Work-life balance, G3 Covid-19 pandemic, G2 Future prospects, G1 Financing of studies, U12 Project group work, U11 Unspec. mention of courses/subjects, U10 Exams and exam preparation, U9 Exercise sheets, U8 Lecture content, U7 Preparing and following up on lectures, U6 Lab reports, U5 Study-related self-regulation, U4 Study routine/time management, U3 Study organization, U2 Transition school-university/ start of semester, U1 Study conditions},
   ytick pos=left,
xtick pos=left,
 legend columns=1, legend cell align = left,legend style = {draw = none},
 legend style = {at ={(0.47,0.91)}, anchor = south west},
 bar width = 5pt,
  ]
\addlegendimage{empty legend}
\addlegendentry{\scalebox{1}[1]{\ref{A1}} Group A1}
\addlegendimage{empty legend}
\addlegendentry{\scalebox{1}[1]{\ref{A2}} Group A2}
\addlegendimage{empty legend}
\addlegendentry{\scalebox{1}[1]{\ref{B1}} Group B1}


\draw[ultra thick] (0,160) -- (400,160);
\draw[ultra thick] (0,80) -- (400,80);
\draw[ultra thick] (0,20) -- (400,20);

\addplot+[blue!60!,area legend,
    draw=black, error bars/.cd, x dir=both, x explicit, error mark options={black,mark size=2pt,line width=.7pt,rotate=90
     },  error bar style={line width=.7pt}
      ] 
		coordinates{
(	4.6, .2)
(	2.6, .4)
(   2.8, .6)
(	7.1, .8)
(	8.3, 1)
(   0.2, 1.2)
(   2.5, 1.4)
(	1.9, 1.6)
(   3.2, 1.8)
(	8.9, 2)
(	10.0, 2.2)
(	16.1, 2.4)
(	2.3,2.6)
(	3.5, 2.8)
(	4.0, 3)
(	6.5, 3.2)
(	6.8, 3.4)
(	2.2, 3.6)
(	2.0, 3.8)
(	4.4, 4)
}; \label{B1}

\addplot+[red!60!,area legend,
    draw=black, error bars/.cd, x dir=both, x explicit, error mark options={black,mark size=2pt,line width=.7pt,rotate=90
     },  error bar style={line width=.7pt}
      ] 
		coordinates{
(	4, .2)
(	4.5, .4)
(   4, .6)
(	2.3, .8)
(	5.4, 1)
(   0.6, 1.2)
(   1.6, 1.4)
(	2.6, 1.6)
(   0, 1.8)
(	14.9, 2)
(	13.7, 2.2)
(	20.6, 2.4)
(	2.6,2.6)
(	2.7, 2.8)
(	6.5, 3)
(	4.8, 3.2)
(	3.6, 3.4)
(	2.0, 3.6)
(	0.9, 3.8)
(	2.7, 4)
}; \label{A2}

\addplot+[black!60!,area legend,
    draw=black, error bars/.cd, x dir=both, x explicit, error mark options={black,mark size=2pt,line width=.7pt,rotate=90
     },  error bar style={line width=.7pt}
      ] 
		coordinates{
(	5.7, .2)
(	3.6, .4)
(   3.8, .6)
(	5.6, .8)
(	6.2, 1)
(   2.9, 1.2)
(   1.5, 1.4)
(	2.5, 1.6)
(   0, 1.8)
(	13.1, 2)
(	9.7, 2.2)
(	14.6, 2.4)
(	4.2,2.6)
(	2.5, 2.8)
(	4.7, 3)
(	6.4, 3.2)
(	5.7, 3.4)
(	2.3, 3.6)
(	1.7, 3.8)
(	3.4, 4)
}; \label{A1}

\end{axis}
\end{tikzpicture}\vspace{-0.3cm}
\caption{Percentage of codings per category across all measuring points for groups~A1 (2,216 codings), A2 (1,248 codings), and B1 (2,359 codings), i.e., cohorts~A and B in their first or second semester.}
\label{fig:stressors}
\end{figure}
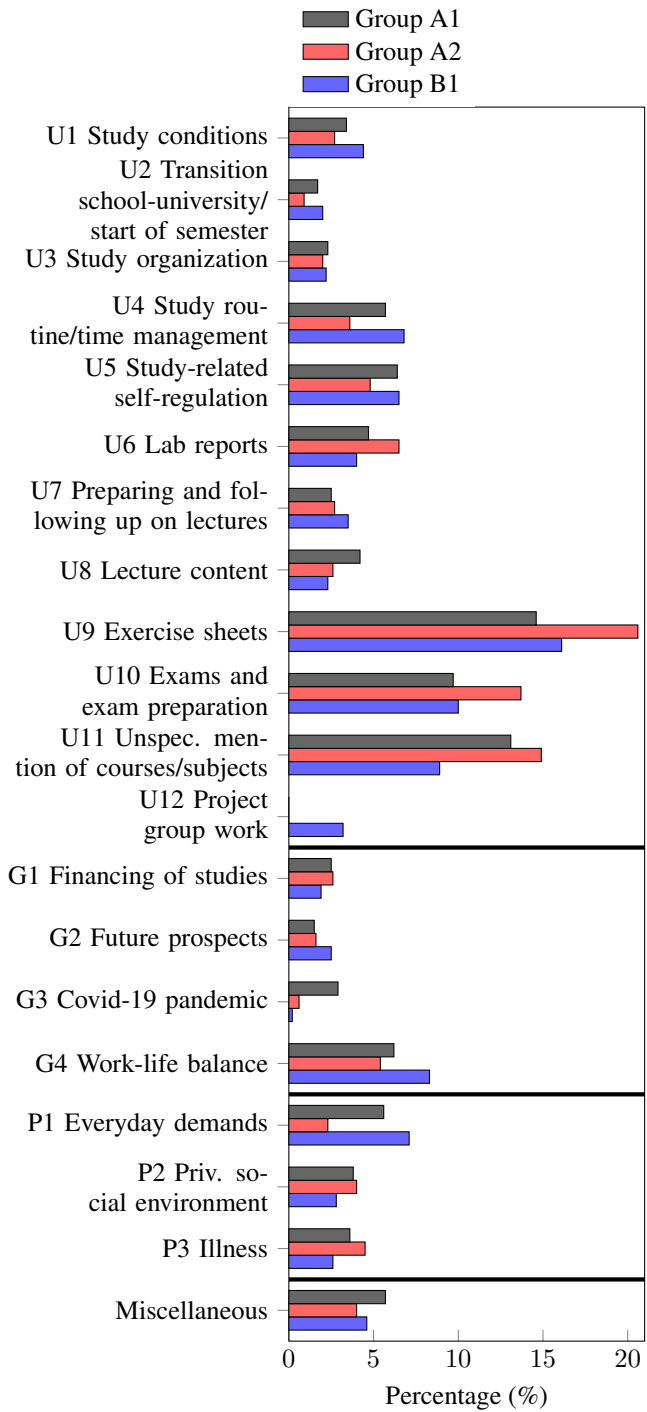

\begin{figure*}[htb]
\centering
\begin{minipage}[t]{0.45\textwidth}
\centering
\textbf{Exercise sheets (U9)}
        \begin{tikzpicture}[scale=.46]
    \pie[
        color={gray, gray!70!, gray!
        40!},
        text=pin,
        explode=0.1
    ]{
        32.7/Math (314),
        9.9/Physics (95),
        57.4/Unspecified (552)
    }
\end{tikzpicture}
\end{minipage}
\hfill
\begin{minipage}[t]{0.5\textwidth}
\centering
\textbf{Unspecific mention of courses (U11)}
        \begin{tikzpicture}[scale=.46]
    \pie[
        color={gray, gray!70!, gray!
        40!,gray!10!},
        text=pin,
        explode=0.1
    ]{
        44.3/Math (304),
        8.7/Physics (60),
        32.3/Physics labs (222),
        14.7/ All other (101)
    }
\end{tikzpicture}
\end{minipage}
\caption{Percentage of codings within \textit{Exercise sheets} (U9) and \textit{Unspecific mention of courses} (U11) in all three groups that can be related to math, physics, or lab courses, as well as all unspecified or other courses. Absolute numbers are given in parentheses.}
\label{fig:Subcategories}
\end{figure*}
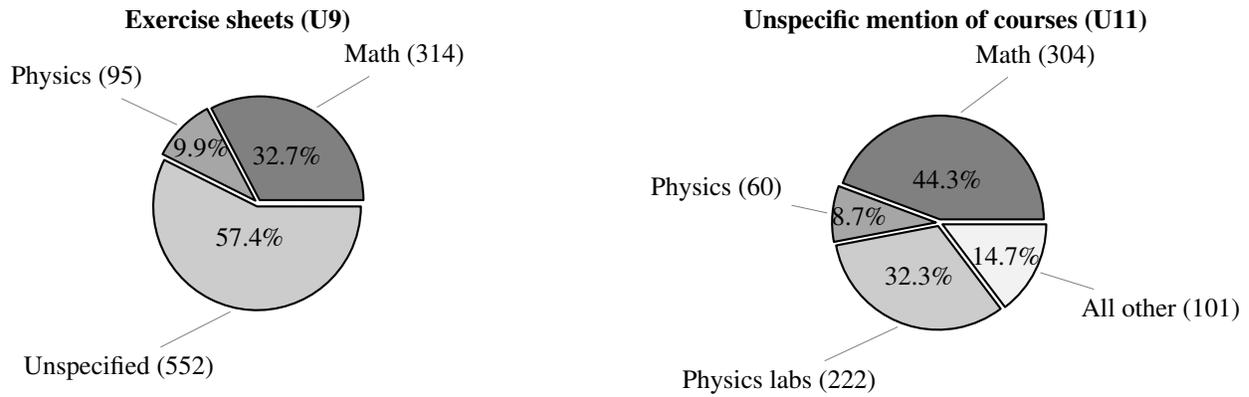

Figure~\ref{fig:Subcategories} and Table~\ref{tab:U9U11} in the appendix provide more detail for the frequent categories U9 and U11 by further subdividing the codings according to the associated courses. Regarding \textit{exercise sheets} (U9), 57\% of all coded responses did not specify the course associated with the exercise sheet. Of the remaining responses, 77\% pertained to math courses, showing significantly less association with physics courses. For groups~A1 and A2, exercise sheets in the course Mathematics for Physicists were predominant, accounting for 85\% and 67\% of all responses with identifiable courses, respectively. In group~B1, exercise sheets for the courses Mathematics for Physicists, Differential and Integral Calculus, and Experimental Physics~I were mentioned with similar frequency, ranging from 31\% to 36\% of all responses with identifiable courses.

Likewise, the \textit{unspecific mention of courses and subjects} (U11) can be further delineated. Across all three groups, 44\% of the codings categorized as U11 were associated with math courses and 32\% with the physics lab course, whereas physics and all other courses had a significantly smaller representation. In groups~A1 and A2, Mathematics for Physicists emerged once more as the most frequently mentioned course (47\% resp. 39\% of all responses), followed by the physics lab course (22\% resp. 33\%); all other courses were mentioned in less than 10\% of the responses. In group~B1, the physics lab course was most frequently listed (45\% of all responses), with other courses (e.g., those for higher years of study or in the second subject for teacher training students) occupying the second position (19\%). Here, Mathematics for Physicists is ranked third, accounting for only 11\% of the responses.

In summary, this shows that the math exercise sheets, the math courses themselves, particularly Mathematics for Physicists, and the physics lab course were the most frequently listed exercise sheets/courses in the codings of U9 and U11.

\subsubsection{Shift in the frequency of the stressors over the semester}

\begin{figure*}
        \centering
    \includegraphics[width=\textwidth]{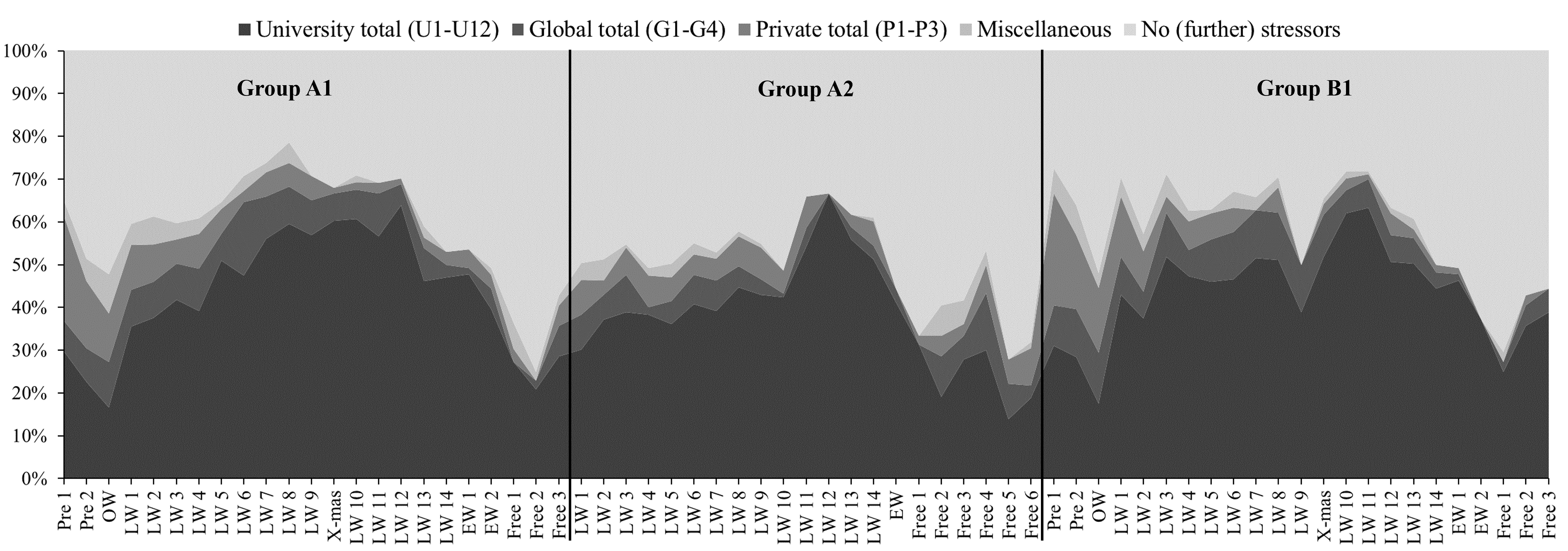}\vspace{-0.3cm}
    \caption{Percentage of stressors listed by the students per each measuring point for all investigated groups, broken down according to the dimensions of the category system (university, global, private, and miscellaneous, cf. Table~\ref{Categories}). The light gray area represents the percentage of open text fields that remained blank when a participating student listed no or fewer than three stressors per measuring point.}
    \label{fig:Dimensions}
\end{figure*}

Figure~\ref{fig:Dimensions} shows how many codings were made in each dimension of the category system for each measuring point for groups A1, A2, and B1. Here, 100\% represents the scenario in which each participant would have named three stressors. Consequently, the light gray area illustrates the percentage of open text fields at each measuring point that were left blank by the students. Thus, the figure shows that the relative number of reported stressors varied throughout the semester. On average, students reported 1.7 out of a maximum of three stressors per measuring point. During the 42 LW measuring points, they reported on average $M=1.9$ ($SD=0.2$) stressors and therefore significantly more than in the twenty-five other measuring points in other phases of the semester with $M=1.4$ ($SD=0.4$) stressors (Welch's t-test, $p<0.001$, $d=1.7$). University-related stressors predominated throughout the entire semester, with 60\% to more than 90\% of all stressors at each measuring point being associated with this dimension. The only exceptions were the pre-course and the orientation week when students were just entering the university and also had to deal with many private and global stressors.

\begin{figure*}[htb]
    \centering
    \includegraphics[width=\textwidth]{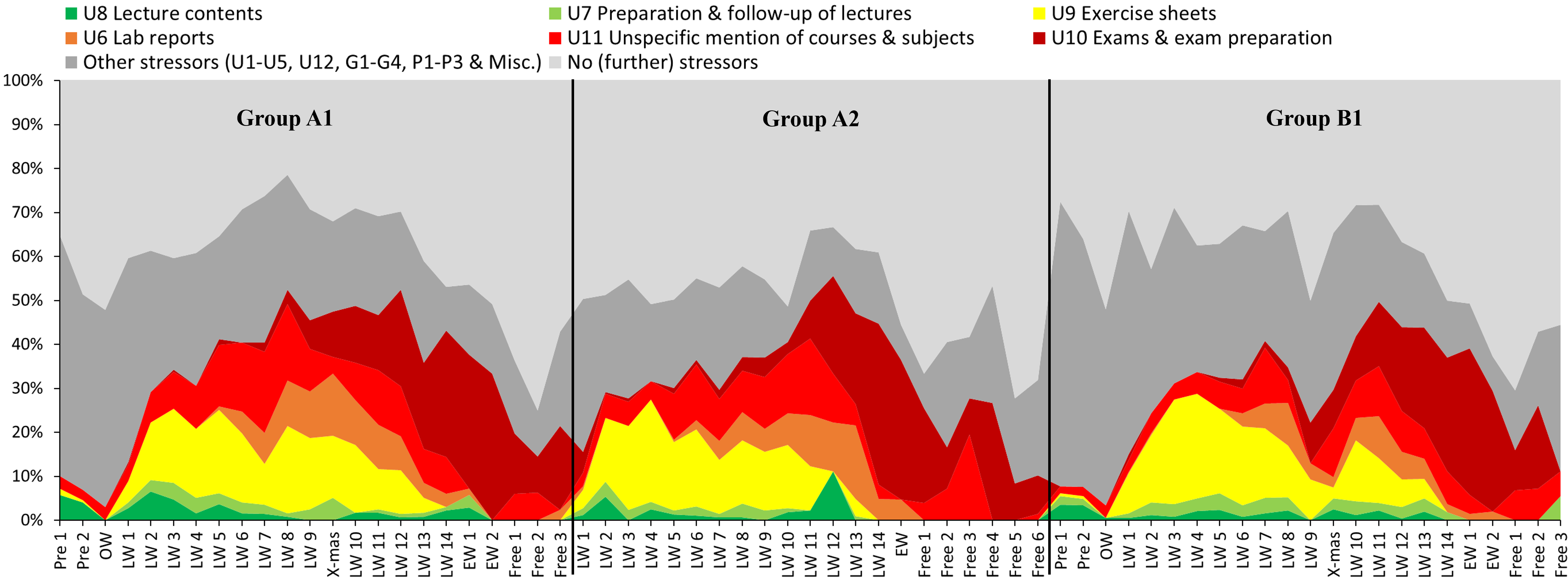}
    \caption{Percentage of stressors listed by students at each measuring point for all groups studied, broken down by specific categories of the university dimension of the category system (cf. Table~\ref{Categories}). In dark gray, all other categories are summarized. The light gray area represents the percentage of open text fields that remained blank when a participating student listed no or fewer than three stressors per measuring point.}\vspace{-0.3cm}
    \label{fig:DevelopmentOfStressors}
\end{figure*}

Figure~\ref{fig:DevelopmentOfStressors} shows the percentage of specific university-related sources of stress mentioned throughout the semester for all three groups. This reveals distinct phases within the semester in which different stressors became more or less significant. Throughout the lecture weeks, the lecture content and the preparing/following up on lectures were frequently mentioned. At the same time, the exercise sheets were also mentioned with a strongly increasing percentage in the first two to three lecture weeks, followed by a fairly steady decline over subsequent weeks. Instead, from the middle of the lecture time, the lab reports and the preparation of exams began to gain significance. The latter became the dominant stressor at the end of the lecture time and during the exam phase. Over the whole lecture time and the lecture-free time thereafter, further courses and subjects were unspecifically mentioned; as described previously, these were mostly mentions of the math courses and the lab course during the semester and specific courses (e.g., a programming course) in the lecture-free time.

\subsubsection{Correlation between perceived stress and number of stressors}\label{correlationstressstressors}

The repeated measures correlation \cite{Bakdash.2017} between the total stress score and the number of mentioned stressors (zero to three) per participant and measuring point across the total data set is $r_{rm}(2887)=0.28$, CI $99\%\;[0.24, 0.33]$, $p<0.001$.

\section{Discussion}

\subsection{Discussion of findings according to RQs}

\subsubsection{Trajectory of perceived stress (RQ1a)}

The trajectory of perceived stress is similar for the first and second semesters and different cohorts of physics students. This might be due to stress perceptions being comparable in the first semester at least on the cohort level and similar study structures of the first and second semesters (e.g., comparable courses and schedules). The significantly lower stress level in period~1 for group~A2 could be explained by the fact that most of the students in group~A1 had already started with the semester due to the preceding pre-course and orientation week, while the students in group~A2 had just started in the new semester during that time as they had a lecture-free and largely exam-free time during the weeks beforehand. The lower stress level during the pre-course and the orientation week can be explained by the different structures and number of tasks and demands during these specific weeks in comparison to the lecture time. The lower stress perceptions in the first weeks of the lecture-free time in comparison to the initial value in the pre-course might be explained by students feeling relieved after exam completion. The slight increase in stress perception in later lecture-free weeks could then be explained by a reduction of this relaxation effect in combination with the second exam phase for some of the students and additional courses like the programming course. During Christmas break, stress perception decreases in group~A1 but not comparably in group~B1. Anecdotal reports by some students from group~B1 indicate that stress perception at that time was caused by the conflict between the desire for vacation and recreation on the one hand, and the list of outstanding study tasks on the other. It is also important to note that only those students who read university e-mails during this vacation time will participate.

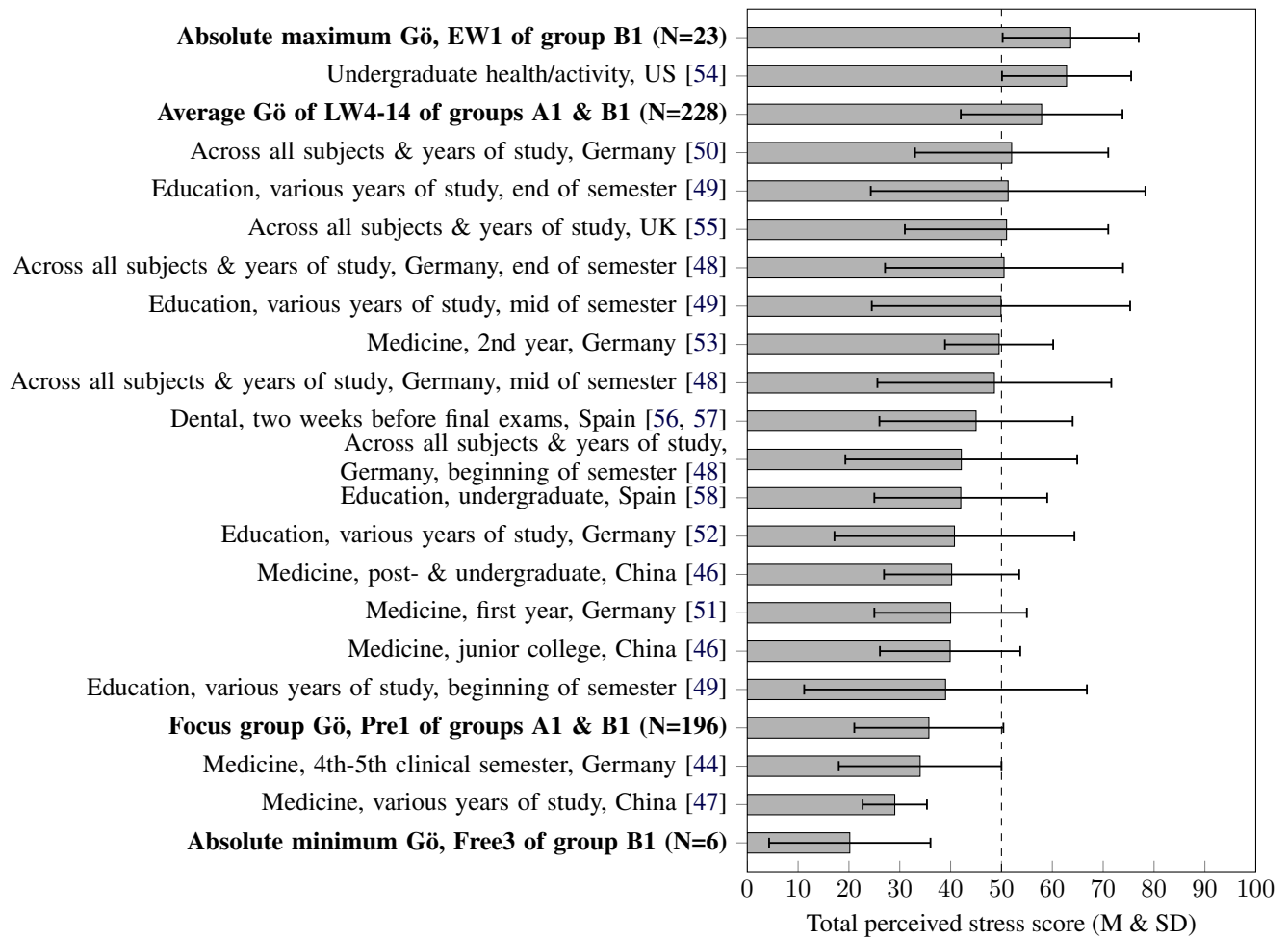
\begin{figure*}[htb]
\flushleft
\begin{tikzpicture}
\begin{axis}[width=.48\textwidth, height=13cm, xbar=3pt,
  ymax=4.35,  
  xmin=0, xmax=100,
  xtick={0,10,20,30,40,50,60,70,80,90,100},
  ymin =-0.15,
  xlabel={Total perceived stress score (M \& SD)},
  ytick = {0,.2,.4,.6,.8,1,1.2,1.4,1.6,1.8,2,2.2,2.4,2.6,2.8,3,3.2,3.4,3.6,3.8,4,4.2},
  yticklabel style={text width=.57\textwidth,align=right, },
  yticklabels={\textbf{Absolute minimum Gö, Free3 of group~B1 (N=6)},{Medicine, various years of study, China} \cite{Jiang.2023},{Medicine, 4$^\text{th}$-5$^\text{th}$  clinical semester, Germany} \cite{Fliege.2001},\textbf{Focus group Gö, Pre1 of groups~A1 \& B1 (N=196)},{Education, various years of study, start of semester} \cite{Obermeier.2021},{Medicine, junior college, China} \cite{Meng.2020},{Medicine, first year, Germany} \cite{Heinen.2017},{Medicine, post- \& undergraduate, China} \cite{Meng.2020},{Education, various years of study, Germany} \cite{Goppert.2021}, {Education, undergraduate, Spain} \cite{MartinezRubio.2023},{Across all subjects \& years of study, Germany, start of semester} \cite{Obermeier.2022},{Dental, two weeks before final exams, Spain} \cite{MonteroMarin.2014,MonteroMarin.2014b},{Across all subjects \& years of study, Germany, mid of semester} \cite{Obermeier.2022},{Medicine, 2$^\text{nd}$ year, Germany} \cite{Moldt.2022},{Education, various years of study, mid of semester} \cite{Obermeier.2021},{Across all subjects \& years of study, Germany, end of semester} \cite{Obermeier.2022},{Across all subjects \& years of study, UK} \cite{vanderFeltzCornelis.2020},{Education, various years of study, end of semester} \cite{Obermeier.2021},{Across all subjects \& years of study, Germany} \cite{Buttner.2013},\textbf{Average Gö of LW4-14 of groups~A1 \& B1 (N=228)},{Undergraduate health/activity, US} \cite{LargoWight.2005},\textbf{Absolute maximum Gö, EW1 of group~B1 (N=23)}},
   ytick pos=left,
xtick pos=left,
 legend columns=1, legend cell align = left,legend style = {draw = none},
 legend style = {at ={(0,1)}, anchor = south west},
 bar width = 8pt,
  ]
\draw[dashed] (50,0.1) -- (50,1000);

\addplot+[gray!60!,area legend,
    draw=black, error bars/.cd, x dir=both, x explicit, error mark options={black,mark size=2pt,line width=.7pt,rotate=90
     },  error bar style={line width=.7pt}
      ] 
		coordinates{
  		(63.61, 4.2)+-(13.38, 4.2)
    (	62.8, 4)+-(12.7,4)
    (57.9, 3.8)+-(15.9,3.8)
    	(	52, 3.6) +-(19,3.6)
      	(    51.3, 3.4) +-(27.00,3.4)
        (	51, 3.2) +-(20,3.2)
        (	50.50, 3) +-(23.40,3)
        (	49.90, 2.8	) +-(25.40,2.8)
    	(	49.53, 2.6) +-(10.64,2.6)
		(	48.60, 2.4	) +-(23.00,2.4)
        (45, 2.2) +- (19, 2.2) 
		(	42.10, 2	) +-(22.80,2)
  		(	42, 1.8	) +-(17,1.8)
		(   40.75, 1.6)+-(23.60,1.6)
		(	40.2, 1.4) +-(13.3,1.4)
		(   40, 1.2)+-(15,1.2)
        (	39.9, 1	) +-(13.8, 1)
        (   39.00, .8)+-(27.80,.8)
        (35.72, .6)+- (14.66,.6)
        (	34, .4) +-(16,.4)
        (	29.019, .2) +-(6.325,.2)
        (20.17, .0)+- (15.87, .0)
}; \label{mean}
\end{axis}
\end{tikzpicture}\vspace{-0.3cm}
\caption{Comparison of selected total perceived stress scores (mean and standard deviation) from the Göttingen data set, highlighted in bold and with the abbreviation Gö, with comparative data from the literature as summarized in Table~\ref{tab:comparison}.}
\label{fig:Comparison}
\end{figure*}

In Figure~\ref{fig:Comparison}, we compare selected measuring points from the Göttingen data set with findings from studies listed in Table~\ref{tab:comparison} that have also used the PSQ with students. These comparisons need to be made with caution as the cited studies used different versions of the PSQ regarding language, number of items, or rating scales, and they were conducted in different countries, in different years, and among varying groups of students. However, the comparison reveals remarkable dynamics in the Göttingen data set since the measuring points with the minimum and maximum total stress scores (Free3 and EW1 for group~B1) under- and over-exceed all stress scores we can find in the literature and differ significantly from each other (Welch's t-test $t(7)=6.16$, $p<0.001$, $d=3.1$). The score of the focus group (Pre1 for groups~A1 and B1) is still very low in comparison to the comparative data. The average of the mean stress score of each participant in groups~A1 and B1 during LW4 and 14 (except Christmas break) is not just much higher than that of the Göttingen focus group (Welch's t-test $t(416)=14.94$, $p<0.001$, $d=1.4$). In fact, it is also higher than most of the comparative scores in the literature, especially the highest score at the end of the semester measured by Ref.~\cite{Obermeier.2022} (Welch's t-test $t(284)=6.48$, $p<0.001$, $d=0.3$), the score which is the most comparable with our study as the same PSQ version was used at a German university at a similar time. Nevertheless, our findings are consistent with Ref.~\cite{Obermeier.2022} in that they also found an increasing stress perception within the first half of the lecture time, with no significant changes thereafter until the end of the lecture time.

In addition to the comparison with data from the literature, the statistically significant differences between the different measuring periods (cf. Sec.~\ref{StressDiffgroups}) and the use of the scheme for data interpretation from Sec.~\ref{datainterpretation} show that the investigated students report, again on average, \textit{slightly increased} stress perceptions during most of the lecture time in comparison to pre-course levels, implying that many students even feel very stressed during that time. Such a persistently high stress level might be considered endangering.

Regarding the two considered predictors, the findings show that gender has no significant effect on stress perceptions in the first pre-course week, while students with a lower high school graduation grade have significantly higher stress perceptions during this time. However, Figure~\ref{fig:ModifiedTrajectorySplit} suggests that this difference disappears during the lecture weeks. This leads to the assumption that if the university does not yet have a greater effect on students' stress perceptions, the predictor becomes relevant in explaining differences in stress perceptions, while during the semester, the university seems to affect students similarly, regardless of high school graduation grades.

\subsubsection{Self-reported workload (RQ1b)}

The workload trajectory is equal for different cohorts and similar for the first and second semesters. Only in period~1, group~A2 indicated a lower workload than group~A1 due to a significantly lower workload for group~A2 at measuring point~LW1. This can again be attributed to the students starting their second semester with a possibly lower workload since the previous week was still lecture-free time, while groups~A1 and B1 already had the pre-course and orientation week. Within a few lecture weeks, the weekly, self-estimated workload reaches a stable plateau of about fifty hours during most lecture weeks, 25\% higher than a typical full-time job. Some students even reported workloads exceeding seventy hours in some weeks. During the Christmas break and lecture period~4 thereafter, the reported workload considerably decreased since students naturally spent less time on their studies during Christmas break. Only during the pre-course and the lecture-free time after the exams was a workload of fewer than forty hours, similar to a full-time job, reported.

The trend in our data follows work by Thiel et al. (2006, as cited in Ref.~\cite{Schulmeister.2011b}) also showing significantly higher workloads during the lecture than the lecture-free time. In absolute figures, our study is similar for the lecture-free time with around twenty hours per week but not for the lecture weeks, when the workload in that study was mostly below the set point of forty hours per week. The comparison with the time budget study by Ref.~\cite{Schulmeister.2011b} also reveals that the workload estimated by the Göttingen students during the lecture time is much higher than one would expect, so the absolute workload figures and the comparison with a full-time job need to be treated with caution here. Possible reasons are various distorting effects that are typical for the self-estimation of workload based on single-item questions: social desirability, conformity pressure, falsifying and distorting the perception of time gaps, recall bias in retrospective surveys, and an unconscious orientation to the well-known amount of forty hours for a typical working week \cite{Schulmeister.2011b}. Further, students may estimate their workloads differently, e.g., by estimating only the actual time spent on their studies or including time for breaks, commuting, etc.

While the comparison with reference data suggests that the Göttingen students might have noticeably overestimated their study-related workloads and the extent of that overestimation is difficult to determine, the dynamics within the estimated workloads between different phases of the semester are remarkable. This leads to the conclusion that workloads are not similarly spread across the semester (lecture vs. lecture-free time). Resulting peak loads may cause additional stress.

\subsubsection{Stressors contributing to stress perception (RQ2)}

The stressors, inductively categorized in Table~\ref{Categories}, can be linked to the state of the research \cite{Albrecht.2011,Thiel.2008,Ortenburger.2013,Gro.2011,Schwedler.2017}. For example, \textit{financing of studies} (G1) and \textit{private social environment} (P2) can similarly be found in Refs.~\cite{Ortenburger.2013,Gro.2011,Thiel.2008}. \textit{Everyday demands} (P1) are comparable to dimensions like household and housing situations in Ref.~\cite{Ortenburger.2013}, and \textit{illness} (P3) was also mentioned by Refs.~\cite{Schwedler.2017,Thiel.2008}. \textit{Future prospects} (G2) indicates that students often associate stress with fear of the future and uncertainty \cite{Ortenburger.2013}. \textit{Work-life balance} (G4) picks up the conflicts between studies and private interests mentioned by Ref.~\cite{Gro.2011} and the area of leisure in Ref.~\cite{Ortenburger.2013}. Many university-related stressors can be linked to the work by Ref.~\cite{Schwedler.2017}, for example \textit{labs} (U6), \textit{exams} (U10), self-study-related categories like \textit{preparing and following up on lectures} (U7), or level of difficulty, which is part of the categories \textit{lecture content} (U8) and \textit{exercise sheets} (U9). The importance of \textit{study conditions} (U1) was highlighted by Refs.~\cite{Albrecht.2011,Gro.2011,Thiel.2008}, and the relevance of time pressure and high quantitative demands (cf. \textit{individual daily study routines/time management}) was highlighted by Refs.~\cite{Ortenburger.2013,Schwedler.2017}. \textit{Study-related self-regulation} (U5) can be compared with the self-concept of ability in Ref.~\cite{Gro.2011}.

The predominance of university-related stressors is probably influenced by our mode of data collection typically within lectures and the instructed context of physics studies. However, the high relevance of these academic stressors should not be overlooked because the number of listed stressors mentioned per participant was significantly higher during the lecture time than in other phases of the semester and correlates with the total stress score. This is also in line with the findings of Ref.~\cite{Ortenburger.2013} that studying is the area of life that is most frequently perceived by students as a strong stressor.

Figure~\ref{fig:stressors} highlights the importance of the two main assessment formats as stressors: the weekly \textit{exercise sheets} (U9) for the acquisition of the exam prerequisite and the \textit{exams and exam preparation} (U11). Reasons for their dominance might be their exam nature, which may be linked to stronger stress perceptions, and the actual related workload since physics students usually spend most of their time during the semester on weekly exercise sheets and later exam preparation. Figure~\ref{fig:Subcategories} shows that math exercise sheets, math courses, and physics lab courses are more frequent stressors than basic physics courses. This is in line with former research \cite{Schwedler.2017,Albrecht.2011,Heublein.2017,Muller.2018,Buschhuter.2016} highlighting the difficulty and concurrent relevance of math for study success among science students, as well as the relevance of lab courses for stress perception \cite{Schwedler.2017}.

The Göttingen students never mentioned the voluntary exercise sheets in the calculus-based course Mathematical Methods in Physics and rarely mentioned the course itself. Instead, they frequently referred to the algebra- and proof-oriented courses Mathematics for Physicists and Differential and Integral Calculus with mandatory exercise sheets. This might be explained by the type of course, calculus- versus algebra- and proof-oriented, as well as the obligation to complete the exercise sheets, both influencing perceptions of the courses. The differences in the frequency of mention of the courses Mathematics for Physicists and Differential and Integral Calculus in the open text responses between groups~A1 and B1 can be attributed to different course selection behaviors.

Although the financial situation is one of the most important stressors for students according to Refs.~\cite{Ortenburger.2013,Gro.2011}, \textit{financing of studies} (G1) was coded rather rarely. A plausible explanation is that the financing of studies is a long-term challenge for many students but may not be perceived as an urgent, acute problem in a specific week compared to other university-related stressors, leading to its less frequent mention in the limited open-text fields in our study.

Figure~\ref{fig:DevelopmentOfStressors} reveals a characteristic progression of the categories \textit{exercise sheets} (U9), \textit{lab reports} (U6), \textit{exams and exam preparation} (U10), and the \textit{unspecific mention of courses and subjects} (U11) over the course of the semester. While sharp peaks and sinks in some parts of that graph are only due to low participation rates at these measuring points, the general trend reflects the structure and conditions of the semester at the University of Göttingen (e.g., weekly exercise sheets throughout the entire lecture time, the start of the lab course only in the middle of the semester, and exams immediately after the lecture time). A shift of stressors during the semester, which was also found by Ref.~\cite{Schwedler.2017}, shows that the students have an awareness of their own stress perceptions and stressors during the semester and that the open text field question is suit-
able to reflect this. However, it must be noted that the students were able to state only up to three most salient stressors each.

\subsection{Integrative discussion and possible implications}

In this study, we used three different measures of stress perceptions. The results and discussion show that all three measures provide a coherent and complementary picture. It is coherent as self-estimated workloads and the number of listed stressors correlate with the total stress score and depict similar perceptions among students. The periods within the semester in which students had increased stress perceptions coincide with those in which they also reported, on average, weekly workloads exceeding those of a typical full-time job. Further, the listed stressors shift over the lecture time while the stress level and workload remain stable. It is also remarkable how similar the three measures are when comparing the three groups, even though the circumstances of each group slightly differ, e.g., by different lecturers, the COVID-19 pandemic for group~A1, or the newly introduced project group work for group~B1. 
The open text responses further reveal that the three groups reported very similar stressors, mainly the exercise sheets, math and lab courses, and exams. These elements might dominate overall stress perceptions, leading to comparable total stress scores and workloads across the groups.

The findings are also complementary as all three measures offer distinct perspectives on the students' stress perceptions. The quantitative data from the PSQ and the workload estimations provide insights into the periods of increased stress across the semester and serve as a benchmark for future measurements. The qualitative data help to explain why the students perceive the physics study entry phase as stressful and therefore serve as a basis for future improvements.

Consequently, this work can be a good starting point for discussions about how to design a study program that is appropriately challenging and educational, but at the same time not overwhelming for students. The results can be used as an empirical basis for the design and implementation of tailored, timely support measures, as well as for discussions about institutional changes, e.g., in the context of reaccreditation procedures. In particular, the quantitative findings show which phases of the semester are particularly challenging for students and could therefore be alleviated and scaffolded by additional support measures. The qualitative findings also show that university-related stressors dominate students' stress perceptions, which can be better addressed by university staff than global or particularly private stressors. The codings indicate which elements of the study program are central and most stressful for students in different phases of the semester, so that the most relevant stressors can be addressed first, and measures can be timed to the phase during the semester when these stressors become dominant. In addition, the outcomes of this study can help raise awareness of study-related stress and stressors and related issues such as mental health among faculty, staff, instructors, and students. Finally, the data set can be used as a benchmark for comparison with other study programs and universities, and to evaluate the effectiveness of newly implemented measures, policies, and institutional changes with follow-up measurements.

\subsection{Limitations}

Data collection was conducted at only one university. The open text field responses showed that some sources of stress, and therefore potentially also the stress trajectory, depend on local conditions, study structures, and teaching practices. So, the extent to which stress perceptions and stressors are specific for certain study programs or more general remains uncertain.

Moreover, the mode of data collection may have influenced the findings. In group~A1, we used paper and online surveys, while later we used only online surveys. Depending on the circumstances in the lectures, data were also not always collected in person on the same day of the week. Additional or substituting online surveys were open for participation for an entire week from one measurement point to the next. Both mechanisms led to elongated and sometimes varying periods of data collection. This might have influenced instantaneous stress perceptions and minded stressors. In that regard, also the course in which the survey was situated may have influenced the study outcomes but the surveyed courses are mentioned less frequently than many other courses in the open text responses coded with categories U9 and U11, so students abstracted their stress perceptions from the surveyed course.

As depicted in Figure~\ref{fig:Trajectory}, the participation rate varied from week to week. While more students participated in the first lecture weeks, the number of participants decreased over the semester and was sometimes influenced by specific incidents like canceled lectures (e.g., LW12 of group~A2). This trend aligns with the decreasing number of students attending the lectures over the semester because particularly students in more advanced semesters often skipped lectures later in the semester. Further influences could have been fatigue in survey participation after so many measurement repetitions or dropout from the entire study program, which could unfortunately not be tracked in the used study design. To encounter the resulting selection bias and gaps in the students' individual trajectories, we used online surveys for a lower participation threshold also for students outside the campus and aggregated measuring periods in the quantitative analysis.

The open text field responses may be influenced by social desirability and various distorting effects related to workload data mentioned before. The limited number of three open text fields for stressors might have resulted in some students not mentioning certain stressors contributing to perceived stress but, e.g., just the most immediate or most recent stressors. This recency bias could skew the data in favor of short-term or immediate stressors over long-term or underlying ones, so the true diversity or intensity of stressors might be underestimated. The shortness of the text fields sometimes also led to imprecise and difficult-to-interpret responses.

\subsection{Outlook: Future perspectives}

Findings and discussed limitations lead to five future perspectives. First, we will extend the investigation of stress perception to other universities and higher years of study in order to characterize and compare different physics study entry phases in terms of resulting stress perceptions.

Second, the category system of stressors will be utilized to develop a standardized questionnaire with a Likert rating scale. This will likely provide more consistent and standardized data on stressors and possible shifts over time.

Third, we will conduct group interviews with first-year and advanced physics students to reflect perceived stressors and existing coping strategies and to discuss possible improvements that could reduce stress perception for future cohorts.

Fourth, we will investigate more parameters related to students' affection and attitudes: their motives for choosing a physics study program \cite{Thiel.2008}, their intentions to change the major or even drop out of university (cf. Ref.~\cite{Baulke.2022,Diederich.2024}), their sense of belonging to the university (cf. Ref.~\cite{Baumert.2008}) and the physics community (cf. Ref.~\cite{Feser.2023c}), their domain-specific growth mindset \cite{Diederich.2024}, their study-related emotions \cite{Pekrun.2005,Breyer.2016}, self-efficacy, and performance characteristics such as prior math and physics skills. These measures will allow a complex, multi-dimensional analysis of relevant factors of the physics study entry phase to better understand and possibly even predict dropout, and to aim at tailor-made support measures.

Fifth, we would like to develop and evaluate supportive measures and study-structural modifications to decrease students' stress perceptions and cope with high dropout rates. A first approach will be a short self-learning course fostering a domain-specific growth mindset \cite{Diederich.2024b}. Other approaches could be linked to strengthening students' resilience \cite{Hofmann.2021}, their sense of belonging, or metacognitive strategies. Refs.~\cite{Hafner.2015,Hafner.2014} have shown that just a two-part management training of two to four hours can significantly reduce perceived tension and facilitate perceived control of time. Ref.~\cite{Buttner.2013} suggests mindfulness-based stress prevention measures, Ref.~\cite{LargoWight.2005} interventions in improving "perceived problem-solving abilities, communication skills, and leadership skills for life success" [p.368]. Ref.~\cite{Gro.2011} emphasizes that it is important to identify and address stressors that are perceived as unnecessary and therefore avoidable. An example would be optimized lecture schedules leading to a more effective day structure with a strict separation between on-campus courses and self-study phases so that there are no larger gaps between on-campus courses.

Overall, the outlined plans aim for a multiperspective analysis of the physics study entry phase and specific, evaluated support measures tailored to guide students more effectively through this particularly challenging transition period. The ultimate goals are to lower stress perception, enhance student success, and, hopefully, reduce study dropout rates.

\section{Conclusion}

In the present study, we investigated the stress perceptions of first-year physics students at a German university in a three-semester panel study. Stress perception was measured mostly weekly for two cohorts of students in their first semester and one cohort additionally in their second semester using the PSQ and open text fields for reporting stressors and self-estimated weekly workloads. The findings show no major differences between the two cohorts in their first semester or between the first and second semesters of the one cohort that was studied in both semesters. Instead, the stress perceptions follow a characteristic trajectory with an increasing level in the first lecture weeks of the semester, reaching a stable level of, on average, \textit{slightly increased} stress in the lecture time and exam phase. This is followed by a decreasing stress level in the lecture-free time afterward. Self-estimated workload shows a similar behavior with a high correlation. The open text fields reveal underlying stressors that are mostly university related and characteristic of certain phases of the semester. Important stressors are the exercise sheets, especially in math courses, the lab course, and lab reports, as well as exam preparation and the exams themselves. The study provides profound and coherent insight into physics students' stress perceptions based on quantitative and qualitative measures, and it also reveals a high stress perception in comparison to reference data. This can serve as a basis for further studies at other universities, for prospective supportive measures for students, and for institutional changes in physics study programs to improve studyability and reduce high dropout rates. 

\section*{Data availability statement}
The data set of this study is openly available as supplementary material. There, the open text field responses are coded according to the presented category system, as the responses are only available in German and are very sensitive. They can be shown to interested readers upon request to the authors.

\section*{Acknowledgments}
We would like to thank all the students in the three courses studied for their repeated and continuous participation in the survey, without which this paper would not have been possible. We would also like to thank our student assistants, Marlene Breither, Stine Gerlach, and Laura Pflügl, for their support in the data curation and interrating process.

\section*{Ethical statement}
The survey was pseudonymized. Data collection and storage was organized in accordance with the General Data Protection Regulation of the European Union and coordinated with the data protection officer of the University of Göttingen. All participants were informed in advance about the aims of the survey and the planned data collection and storage and agreed to participate voluntarily. In the case of online surveys, in addition to a time stamp, only personal data that was explicitly requested during the surveys was collected and stored.

\section*{Author contributions according to CRediT}
\footnotesize{\textbf{J.~O.~C.}: data curation (supporting), formal analysis (supporting), validation (equal), visualization (supporting), writing - review \& editing (supporting); \textbf{J.~H.}: writing - review \& editing (supporting); \textbf{L.~H.}: conceptualization (equal), data curation (supporting), investigation (supporting), methodology (equal), visualization (supporting); \textbf{P.~K.}: conceptualization (equal), investigation (supporting), methodology (equal), supervision (lead), writing - review \& editing (supporting), project administration, resources; \textbf{S.~Z.~L.}: conceptualization (equal), data curation (lead), formal analysis (lead), investigation (lead), methodology (equal), validation (equal), visualization (lead), writing - original draft (lead), writing - review \& editing (lead); \textbf{J.~N.}: formal analysis (lead), visualization (supporting), writing - original draft (supporting), writing - review \& editing (supporting); \textbf{S.~S.}: supervision (supporting), resources.}

\newpage
\onecolumngrid
\appendix*
\section{}

\begin{table*}[htb]
\caption{Overview of the questionnaire used for each measuring point (demographic data only once at the beginning of each semester).}
\begin{ruledtabular}
\footnotesize
\begin{tabular}{p{.14\textwidth}p{.41\textwidth}p{.42\textwidth}}
Questionnaire/Item & Specification/wording (in English) & Specification/wording (in German)\\\hline
Demographic data&Gender, field of study, current semester, number of university courses currently attended, specification of the courses attended, assessment of own physics performance, assessment of own maths performance, grade of high school diploma & Geschlecht, Studiengang, Fachsemester, Anzahl aktuell besuchter universitärer Veranstaltungen, Spezifizierung der besuchten Veranstaltungen, Selbsteinschätzung der Physikleistung, Selbsteinschätzung der Mathematikleistung, Abiturnote\\
Perceived Stress Questionnaire (PSQ20) &\textit{In the following you will ﬁnd a series of statements. Please read each one and rate the frequency with which this statement has been true in your life in the last week (on a scale from 1 = "almost never true" to 6 = "mostly true"). Think about your studies! For each statement, please check the box that best applies. There are no right or wrong answers. Evaluate the answers ad hoc, without thinking too long, and do not leave out any questions}.\newline 20 items, modified after Ref.~\cite{Fliege.2001} into a six-point scale (\textit{almost never} (1), \textit{mostly} (6), the gradations in between were not specified) & 
\textit{Im Folgenden ﬁnden Sie eine Reihe von Aussagen. Bitte lesen Sie jede durch und beurteilen Sie die Häuﬁgkeit mit der diese Aussage in der letzten Woche in ihrem Leben zutreﬀend war (auf einer Skala von 1 = "triﬀt fast nie zu" bis 6 = "triﬀt meistens zu"). Denken Sie dabei an Ihr Studium! Kreuzen Sie bitte bei jeder Aussage das Feld an, das am besten zutriﬀt. Es gibt keine richtigen oder falschen Antworten. Beurteilen Sie die Antworten ad hoc, ohne lange nachzudenken, und lassen Sie keine Frage aus.}\newline
20 Items von Ref.~\cite{Fliege.2001}, modifiziert durch eine sechsstufige Skala (\textit{fast nie} (1), \textit{meistens} (6), die Zwischenstufen blieben unbenannt)\\
Self-estimated workload& \textit{Estimate the total amount of time you spent on your studies within the last week (lecture, tutorial, self-study, etc.) in
hours.}&\textit{Schätzen Sie den zeitlichen Aufwand, den Sie innerhalb der letzten Woche für Ihr Studium ins\-ge\-samt aufgewendet haben (Vorlesung, Übung, Selbststudium, etc.) in Stunden.}\\
Specification of stressors& \textit{Specify up
to three causes that are currently generating a load (strongest load first). Be as precise as possible in your indication of what exactly the load consists of.} &\textit{Geben Sie bis zu drei Ursachen an, die gerade eine Belastung erzeugen (die stärkste Belastung zuerst). Seien Sie dabei möglichst präzise in Ihrer Angabe, worin genau die Belastung besteht.}
\end{tabular}
\end{ruledtabular}
\label{tab:instrument}
\end{table*}

\begin{figure*}[h!]
    \centering
    \includegraphics[width=.8\textwidth]{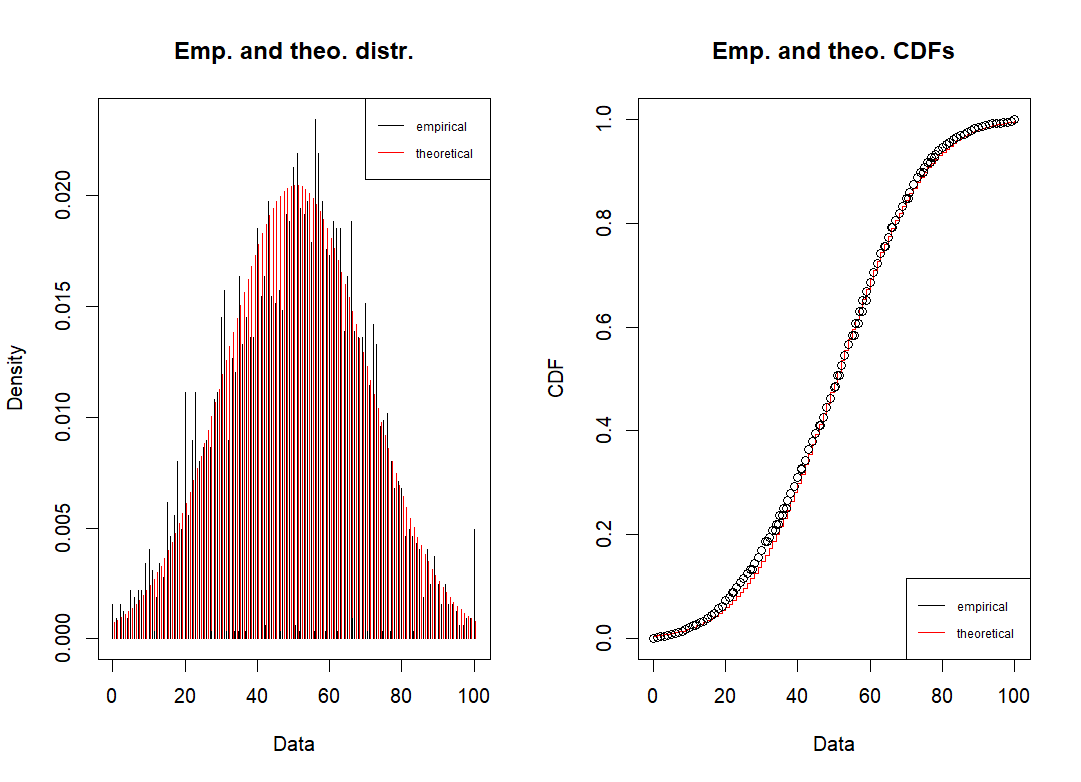}\vspace{-0.3cm}
    \caption{On the left is the empirical and theoretical density distribution of total perceived stress scores for the full data set ($N=3,241$). On the right is the corresponding cumulative distribution function (CDF). Visual inspection suggests that the total perceived stress scores are nominally distributed.}
    \label{fig:normaldistribution}
\end{figure*}

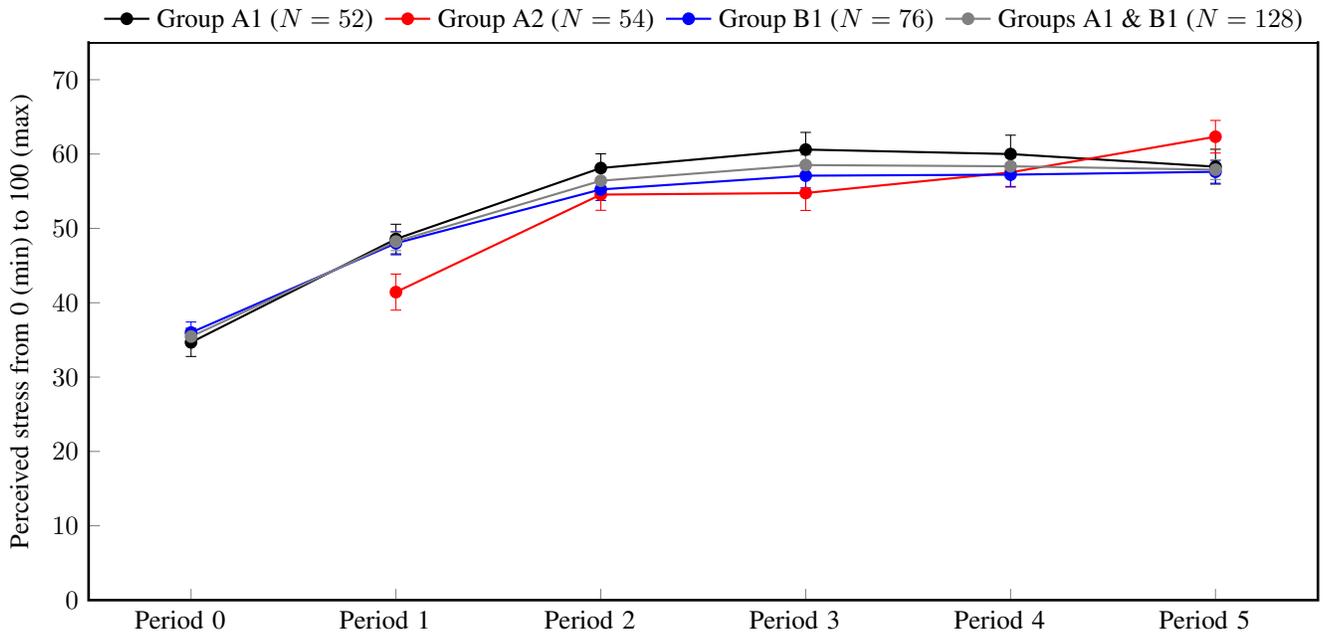
\begin{figure*}[htb]
\centering
\begin{tikzpicture}
\begin{axis}[ width=\textwidth, height=9cm, 
grid=none, line width= 1pt,
  ymax=75,  
  xmin=0.5, xmax=6.5,
  ytick={0,10,20,30,40,50,60,70,80,90,100},
  ymin =0, 
  ylabel={Perceived stress from 0 (min) to 100 (max)},
  xtick = {1,2,3,4,5,6},
  xticklabel style={text width=1.5cm, },
  xticklabels={Period 0, Period 1, Period 2, Period 3, Period 4, Period 5},
   xtick pos=left,
ytick pos=left,
 legend columns=4, legend cell align = left,legend style = {draw = none},
 legend style = {at ={(0,1)}, anchor = south west},
 bar width = 5pt,
  ]
    \addlegendimage{empty legend}\addlegendentry{\scalebox{1}[1]{\ref{StressA1mod}} Group A1 ($N=52$)
    }
    \addlegendimage{empty legend}\addlegendentry{\scalebox{1}[1]{\ref{StressA2mod}} Group A2 ($N=54$)
    }
    \addlegendimage{empty legend}\addlegendentry{\scalebox{1}[1]{\ref{StressB1mod}} Group B1 ($N=76$)
    }
    \addlegendimage{empty legend}\addlegendentry{\scalebox{1}[1]{\ref{StressA1B1mod}} Groups A1 \& B1 ($N=128$)
    }

\addplot[scatter/classes={a={black}},
    scatter, visualization depends on=\thisrow{ey} \as \myshift,
    every node near coord/.append style = {shift={(axis direction
    cs:0,\myshift)}},
    scatter src=explicit symbolic,
    ]
    plot [mark=*,draw=black, thick, mark options = {solid}, error bars/.cd, y dir = both, y explicit,  error mark options={solid, black, rotate=90,mark size=2pt,}, error bar style ={solid}]
    table[meta=class, x=x, y=y, y error=ey]{
        x   y   ey    class label
        1   34.66882275  1.897749013 1
       2	48.56933198 1.990150664 2
       3	58.11808367 1.913052286 3
       4   60.60300883 2.309923055 4
       5	59.99796683 2.53960722 5
       6 58.3021978 2.342724866 6
        }; \label{StressA1mod}
       
\addplot+[scatter/classes={a={red}},
    scatter, visualization depends on=\thisrow{ey} \as \myshift,
    every node near coord/.append style = {shift={(axis direction
    cs:0,\myshift)}},
    scatter src=explicit symbolic,
    ]
    plot [mark=*,draw=red, thick, mark options = {solid}, error bars/.cd, y dir = both, y explicit,  error mark options={solid, red, rotate=90,mark size=2pt,}, error bar style ={solid}]
    table[meta=class, x=x, y=y, y error=ey]{
        x   y   ey    class label
        2 41.42971236 2.424868219 2
        3 54.54615758 2.112342932 3
        4 54.75776144 2.341057733 4
        5 57.53243767 1.986663993 5
        6 62.32897603 2.191639812 6
        }; \label{StressA2mod}

\addplot[scatter/classes={a={blue}},
    scatter, visualization depends on=\thisrow{ey} \as \myshift,
    every node near coord/.append style = {shift={(axis direction
    cs:0,\myshift)}},
    scatter src=explicit symbolic,
    ]
    plot [mark=*,draw=blue, thick, mark options = {solid, fill=blue}, error bars/.cd, y dir = both, y explicit,  error mark options={solid, blue, rotate=90,mark size=2pt,}, error bar style ={solid}]
    table[meta=class, x=x, y=y, y error=ey]{
        x   y   ey    class label
        1 35.99078947 1.417618545 1
        2 48.00219298 1.554097563 2
        3 55.23917592 1.484995649 3
        4 57.09170129 1.645097724 4
        5 57.22807018 1.58910411 5
        6 57.6152746 1.542410417 ^5
        }; \label{StressB1mod}

        \addplot[scatter/classes={a={gray}},
    scatter, visualization depends on=\thisrow{ey} \as \myshift,
    every node near coord/.append style = {shift={(axis direction
    cs:0,\myshift)}},
    scatter src=explicit symbolic,
    ]
    plot [mark=*,draw=gray, thick, mark options = {solid, fill=gray}, error bars/.cd, y dir = both, y explicit,  error mark options={solid, gray, rotate=90,mark size=2pt,}, error bar style ={solid}]
    table[meta=class, x=x, y=y, y error=ey]{
        x   y   ey    class label
        1 35.45374049 1.138208961 1
        2 48.2325932 1.222141732 2
        3 56.40873219 1.177297964 3
        4 58.51816998 1.357513091 4
        5 58.35334069 1.397216079 5
        6 57.89433715 1.315465093 6
        }; \label{StressA1B1mod}
\end{axis}
\end{tikzpicture}\vspace{-0.3cm}
\caption{Visualization of the derived perceived stress scores (means and standard errors) used for quantitative data analysis after selecting specific measuring points (Pre1, Pre 2, OW, and LW1-14), combining them into six measuring periods, and imputing missing data as described in Sec.~\ref{modificquandata}. Data are presented for all three groups, as well as for groups~A1 and B1, both first semester, combined, as they were treated together when the temporal evaluation of perceived stress over the semester was analyzed (cf. Sec.~\ref{stressdiffperiods}).}
\label{fig:ModifiedTrajectory}
\end{figure*}

\begin{figure*}[htb]
\centering
\begin{tikzpicture}
\begin{axis}[ width=\textwidth, height=9cm, 
grid=none, line width= 1pt,
  ymax=75,  
  xmin=0.5, xmax=6.5,
  ytick={0,10,20,30,40,50,60,70,80,90,100},
  ymin =0, 
  ylabel={Perceived stress from 0 (min) to 100 (max)},
  xtick = {1,2,3,4,5,6},
  xticklabel style={text width=1.5cm, },
  xticklabels={Period 0, Period 1, Period 2, Period 3, Period 4, Period 5},
   xtick pos=left,
ytick pos=left,
 legend columns=4, legend cell align = left,legend style = {draw = none},
 legend style = {at ={(0,1)}, anchor = south west},
 bar width = 5pt,
  ]
    \addlegendimage{empty legend}\addlegendentry{\scalebox{1}[1]{\ref{best}} Best high school graduation grades ($N=30$)
    }
    \addlegendimage{empty legend}\addlegendentry{\scalebox{1}[1]{\ref{lowest}} Lowest high school graduation grades ($N=30$)
    }

\addplot[scatter/classes={a={black}},
    scatter, visualization depends on=\thisrow{ey} \as \myshift,
    every node near coord/.append style = {shift={(axis direction
    cs:0,\myshift)}},
    scatter src=explicit symbolic,
    ]
    plot [mark=*,draw=black, thick, mark options = {solid}, error bars/.cd, y dir = both, y explicit,  error mark options={solid, black, rotate=90,mark size=2pt,}, error bar style ={solid}]
    table[meta=class, x=x, y=y, y error=ey]{
        x   y   ey    class label
        1   33.66087719 2.156041215 1
       2	48.10204678 2.845781491 2
       3	56.22222222 2.412125252 3
       4  60.53646996 2.865218596 4
       5	56.9371345 3.187094091 5
       6 56.66884058 2.756175469 6
        }; \label{best}
       
\addplot+[scatter/classes={a={red}},
    scatter, visualization depends on=\thisrow{ey} \as \myshift,
    every node near coord/.append style = {shift={(axis direction
    cs:0,\myshift)}},
    scatter src=explicit symbolic,
    ]
    plot [mark=*,draw=red, thick, mark options = {solid}, error bars/.cd, y dir = both, y explicit,  error mark options={solid, red, rotate=90,mark size=2pt,}, error bar style ={solid}]
    table[meta=class, x=x, y=y, y error=ey]{
        x   y   ey    class label
        1   39.57075475 2.748453774 1
        2 47.74912281 2.247494911 2
        3 56.1502924 2.506850046 3
        4 56.74274322 2.833127383 4
        5 56.05 3.114227537 5
        6 58.77455388 2.954211459 6
        }; \label{lowest}
\end{axis}
\end{tikzpicture}\vspace{-0.3cm}
\caption{Visualization of the derived perceived stress scores (means and standard errors) of a subset of students from groups~A1 and B1, following the data processing rule described in Sec.~\ref{modificquandata}, who were already presented in the previous Figure~\ref{fig:ModifiedTrajectory} and for whom the high school graduation grade was available. Data are presented for the quartiles of students with the best ($M=1.03$, $SD=0.05$) and the lowest ($M=2.17$, $SD=0.31$) high school graduation grades.}
\label{fig:ModifiedTrajectorySplit}
\end{figure*}
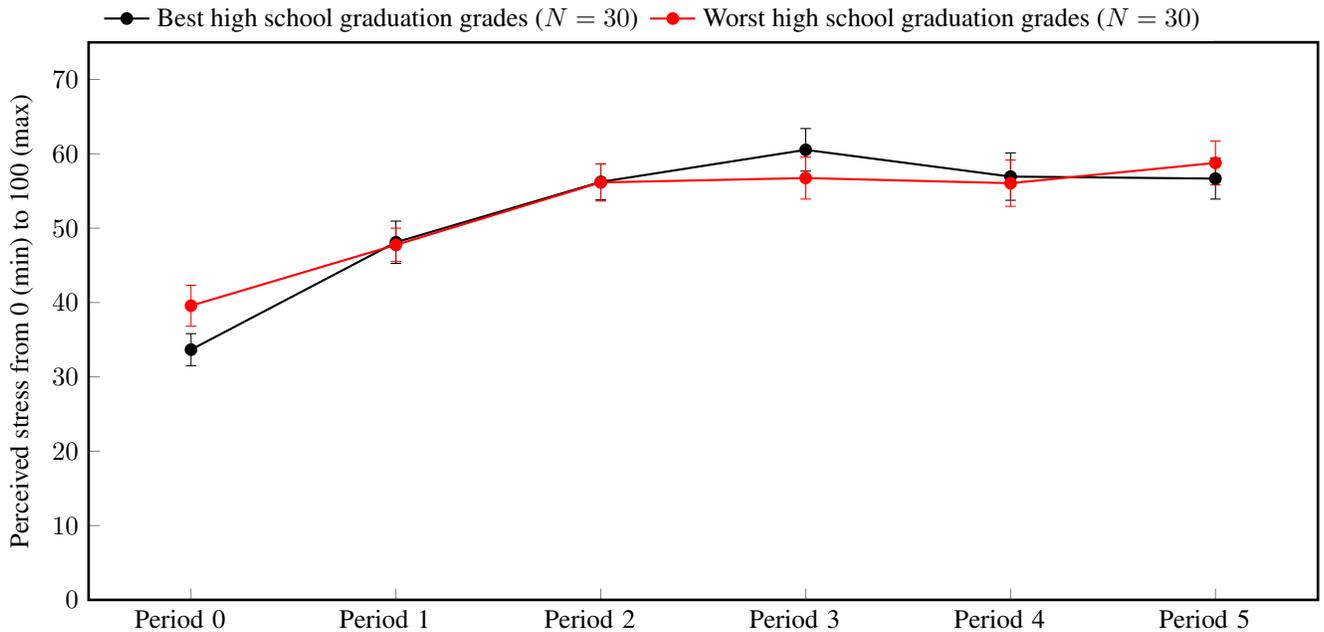

\begin{figure*}[htb]
\centering
\begin{tikzpicture}
\begin{axis}[ width=\textwidth, height=9cm, 
grid=none, line width= 1pt,
  ymax=65,  
  xmin=0.5, xmax=6.5,
  ytick={0,10,20,30,40,50,60},
  ymin =0, 
  ylabel={Self-estimated, weekly study-related workload (in h)},
  xtick = {1,2,3,4,5,6},
  xticklabel style={text width=1.5cm, },
  xticklabels={Period 0, Period 1, Period 2, Period 3, Period 4, Period 5},
   xtick pos=left,
ytick pos=left,
 legend columns=4, legend cell align = left,legend style = {draw = none},
 legend style = {at ={(0,1)}, anchor = south west},
 bar width = 5pt,
  ]
    \addlegendimage{empty legend}\addlegendentry{\scalebox{1}[1]{\ref{WorkmodA1}} Group A1 ($N=52$)
    }
    \addlegendimage{empty legend}\addlegendentry{\scalebox{1}[1]{\ref{WorkmodA2}} Group A2 ($N=54$)
    }
    \addlegendimage{empty legend}\addlegendentry{\scalebox{1}[1]{\ref{WorkmodB1}} Group B1 ($N=76$)
    }
    \addlegendimage{empty legend}\addlegendentry{\scalebox{1}[1]{\ref{WorkmodA1B1}} Groups A1 \& B1 ($N=128$)
    }

\addplot[scatter/classes={a={black}},
    scatter, visualization depends on=\thisrow{ey} \as \myshift,
    every node near coord/.append style = {shift={(axis direction
    cs:0,\myshift)}},
    scatter src=explicit symbolic,
    ]
    plot [mark=*,draw=black, thick, mark options = {solid}, error bars/.cd, y dir = both, y explicit,  error mark options={solid, black, rotate=90,mark size=2pt,}, error bar style ={solid}]
    table[meta=class, x=x, y=y, y error=ey]{
        x   y   ey    class label
        1   18.56457666  0.80752095 1
       2	36.72676282 1.316347375 2
       3	51.70737179 1.326016043 3
       4  54.64865967 1.444391608 4
       5	50.04216149 1.503233816 5
       6 53.40693681 1.222371587 6
        }; \label{WorkmodA1}
       
\addplot+[scatter/classes={a={red}},
    scatter, visualization depends on=\thisrow{ey} \as \myshift,
    every node near coord/.append style = {shift={(axis direction
    cs:0,\myshift)}},
    scatter src=explicit symbolic,
    ]
    plot [mark=*,draw=red, thick, mark options = {solid}, error bars/.cd, y dir = both, y explicit,  error mark options={solid, red, rotate=90,mark size=2pt,}, error bar style ={solid}]
    table[meta=class, x=x, y=y, y error=ey]{
        x   y   ey    class label
        2 28.43774776 1.616786754 2
        3 49.40140796 1.671107134 3
        4 49.50867829 1.670838859 4
        5 51.20518337 1.337850051 5
        6 53.51320806 1.484094959 6
        }; \label{WorkmodA2}

\addplot[scatter/classes={a={blue}},
    scatter, visualization depends on=\thisrow{ey} \as \myshift,
    every node near coord/.append style = {shift={(axis direction
    cs:0,\myshift)}},
    scatter src=explicit symbolic,
    ]
    plot [mark=*,draw=blue, thick, mark options = {solid, fill=blue}, error bars/.cd, y dir = both, y explicit,  error mark options={solid, blue, rotate=90,mark size=2pt,}, error bar style ={solid}]
    table[meta=class, x=x, y=y, y error=ey]{
        x   y   ey    class label
        1 17.5860354 0.950076259 1
        2 30.48815789 1.231212964 2
        3 50.99076131 1.176614388 3
        4 52.77445464 1.384022987 4
        5 49.16666667 1.421356187 5
        6 52.35898169 1.142970981 6
        }; \label{WorkmodB1}

        \addplot[scatter/classes={a={gray}},
    scatter, visualization depends on=\thisrow{ey} \as \myshift,
    every node near coord/.append style = {shift={(axis direction
    cs:0,\myshift)}},
    scatter src=explicit symbolic,
    ]
    plot [mark=*,draw=gray, thick, mark options = {solid, fill=gray}, error bars/.cd, y dir = both, y explicit,  error mark options={solid, gray, rotate=90,mark size=2pt,}, error bar style ={solid}]
    table[meta=class, x=x, y=y, y error=ey]{
        x   y   ey    class label
        1 17.98356779 0.651692931 1
        2 33.02259115 0.942408887 2
        3 51.28188432 0.879352159 3
        4 53.53585043 1.009297794 4
        5 49.52233644 1.038501729 5
        6 52.78471346 0.838998043 6
        }; \label{WorkmodA1B1}
\end{axis}
\end{tikzpicture}\vspace{-0.3cm}
\caption{Visualization of the derived workload (means and standard errors) used for the quantitative data analysis after selecting specific measuring points (Pre1, Pre 2, OW, and LW1-14), combining them into six measuring periods, winsorizing, and imputing missing data as described in Sec.~\ref{modificquandata}. Data are presented for all three groups, as well as for groups~A1 and B1, both first semester, combined, as they were treated together when the temporal evaluation of perceived stress over the semester was analyzed (cf. Sec.~\ref{workloaddiffperiods}).}
\label{fig:Modifiedworkload}
\end{figure*}
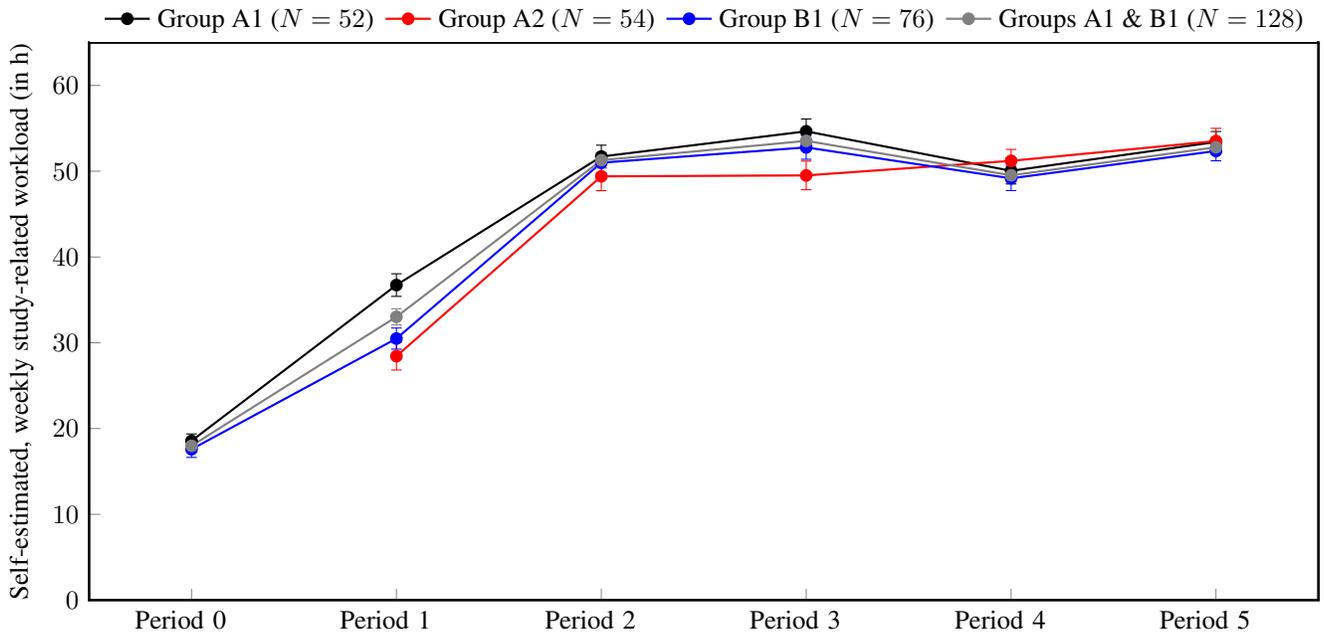

\begin{table*}[htb]
\caption{Bonferroni-corrected p-values of the post-hoc analysis of differences in stress perception and workload between different periods of the semester, separately for the first (groups A1 \& B1) and second (group A2) semesters.}
\footnotesize
\begin{ruledtabular}
\begin{tabular}{p{.086\columnwidth}p{.086\columnwidth}p{.086\columnwidth}p{.086\columnwidth}p{.086\columnwidth}p{.086\columnwidth}p{.086\columnwidth}p{.086\columnwidth}p{.086\columnwidth}p{.086\columnwidth}p{.086\columnwidth}}
&\multicolumn{5}{l}{Stress perception (cf. Sec.~\ref{stressdiffperiods})}&\multicolumn{5}{l}{Workload (cf. Sec.~\ref{workloaddiffperiods})}\\
&Period~0&Period~1&Period~2&Period~3&Period~4&Period~0&Period~1&Period~2&Period~3&Period~4\\\hline
\multicolumn{11}{l}{Groups A1 \& B1}\\\hline
Period~1&<0.001&&&&&<0.001&&&&\\
Period~2&<0.001&<0.001&&&&<0.001&<0.001&&&\\
Period~3&<0.001&<0.001&0.77&&&<0.001&<0.001&0.069&&\\
Period~4&<0.001&<0.001&0.92&1.00&&<0.001&<0.001&1.00&0.002&\\
Period~5&<0.001&<0.001&1.00&1.00&1.00&<0.001&<0.001&1.00&1.00&0.002\\\hline
\multicolumn{11}{l}{Group A2}\\\hline
Period~2&&<0.001&&&&&<0.001&&&\\
Period~3&&<0.001&1.00&&&&<0.001&1.00&&\\
Period~4&&<0.001&0.25&1.00&&&<0.001&1.00&1.00&\\
Period~5&&<0.001&<0.001&<0.001&0.052&&<0.001&0.024&0.030&0.64\\
\end{tabular}
\end{ruledtabular}
\label{tab:pvalues}
\end{table*}

\begin{table*}[htb]
\caption{Percentage (\%) of codings within the categories U9 Exercise sheets and U11 Unspecific mention of courses \& subjects divided by the different courses in the first and second semesters. Data are presented for all three groups~A1, A2, and B1, and for the total sample. The first row indicates the number of codings for which the percentages in each column are calculated. Gaps indicate that the corresponding course could not be coded for the specific category/group because the course was not offered for that group or did not have an exercise sheet.}
\footnotesize
\begin{ruledtabular}
\begin{tabular}{p{.24\textwidth}p{.09\textwidth}p{.09\textwidth}p{.09\textwidth}p{.09\textwidth}p{.09\textwidth}p{.09\textwidth}p{.09\textwidth}p{.09\textwidth}}
&\multicolumn{4}{l}{U9 Exercise sheets}&\multicolumn{4}{l}{U11 Unspecific mention of courses \& subjects}\\
&Group~A1&Group~A2&Group~B1&Total&Group~A1&Group~A2&Group~B1&Total\\\hline
\diagbox[width=.24\textwidth]{Percentage}{N codings}&167&136&249&552&29&17&41&87\\\hline
Experimental Physics I/II&4.6&8.9&11.3&8.4&7.9&2.7&3.3&5.1\\
Mathematics for Physicists I/II&41.4&31.5&10.8&26.6&46.9&38.7&11.4&33.8\\
Differential and integral calculus I/II&2.5&1.2&12.4&6.0&2.1&5.9&6.6&4.5\\
Mathematical Methods in Physics&0.0&&0.0&0.0&4.5&&4.7&3.3\\
Analytical Mechanics&&5.4&&1.5&&8.6&&2.3\\
Physics lab course&&&&&22.1&33.3&44.5&32.0\\
Precourse&&&&&3.4&&1.9&2.0\\
Unspecified&51.5&52.9&65.5&57.4&&&&\\
Unspecific math course&&&&&2.8&0.5&4.3&2.6\\
Unspecific physics course&&&&&0.3&0.0&3.8&1.3\\
Other courses&&&&&10.0&9.1&19.4&12.7\\
\end{tabular}
\end{ruledtabular}
\label{tab:U9U11}
\end{table*}

\FloatBarrier
\twocolumngrid
\bibliography{bibtex1}

\end{document}